%% file: VisualAnalyticsSurvey.tex
\RequirePackage{fix-cm}
\documentclass[twocolumn,compsoc]{cvm}

\usepackage{multirow}
\usepackage{longtable}
\usepackage{booktabs, makecell}
\usepackage{hyperref}

\graphicspath{{figures/},{figures/cvm/}}

\def\ManuscriptInfo{Manuscript received: xxxx-xx-xx; accepted: xxxx-xx-xx.}

\def\PaperTitle{A Survey of Visual Analytics Techniques\\\ for Machine Learning}

\def\AuthorNames{Jun Yuan$^1$, Changjian Chen$^1$, Weikai Yang$^1$, Mengchen Liu$^2$, Jiazhi Xia$^3$, Shixia Liu$^1$(\rm{\Letter})}

\def\AuthorAdress{$1\quad $ & BNRist,  Tsinghua University, Beijing 100086, China. Email: J. Yuan, C. Chen, W. Yang, \{yuanj19, ccj17, ywk19\}@mails.tsinghua.edu.cn; S. Liu, shixia@tsinghua.edu.cn(\rm{\Letter}).\\
    $2\quad $ & Microsoft, Redmond 98052, United States, E-mail: mengcliu@microsoft.com.\\
    $3\quad $ & Central South University, Changsha 410083, China, E-mail: xiajiazhi@csu.edu.cn.\\
}

\def\firstheader{DOI 10.1007/s41095-xxx-xxxx-x\hfill Vol. x, No. x, month year, xx--xx\\[5mm]~Article Type (Research/Review)}

\headevenname{{J. Yuan, C. Chen, W. Yang \etal} }

\newcommand{\changjian}[1]{\textcolor{black}{#1}}
\newcommand{\jiazhi}[1]{\textcolor{black}{#1}}
\definecolor{mygreen}{RGB}{116, 196, 118}

\newcommand{\yuanjun}[1]{\textcolor{black}{#1}}
\newcommand{\mengchen}[1]{\textcolor{black}{#1}}
\newcommand{\revision}[1]{\textcolor{black}{#1}}

\def \etal {{\emph{et al}.\thinspace}}
\def \eg {{\emph{e.g}.\thinspace}}
\def \ie {{\emph{i.e}.\thinspace}}

\begin{document}

\MakePageStyle

\input{0-abstract.tex}
\input{1-introduction.tex}
\input{2-scope.tex}

\input{3-before.tex}
\input{4-in.tex}
\input{5-after.tex}
\input{6-opportunity}
\input{7-conclusion}



\bibliographystyle{CVM}

{\normalsize  \bibliography{ref}}

\Author{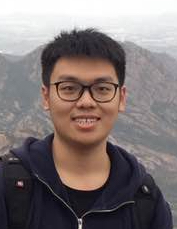}{Jun Yuan}
{is currently a Ph.D. student at Tsinghua University. His research interests are in explainable artificial intelligence. He received a B.S. degree from Tsinghua University. \\ \\ \\}

\Author{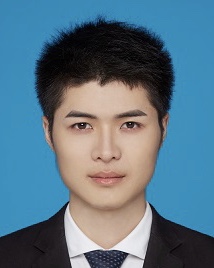}{Changjian Chen}
{is now a Ph.D. student at Tsinghua University. His research interests are in interactive machine learning. He received a B.S. degree from the University of Science and Technology of China.}

\Author{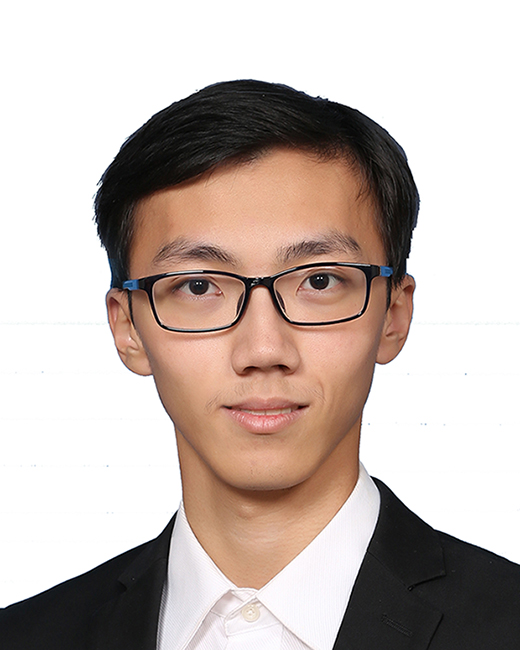}{Weikai Yang}
{is a graduate student at Tsinghua University. His research interest is in visual text analytics. He received a B.S. degree from Tsinghua University. \\ \\ \\ \\ \\}

\Author{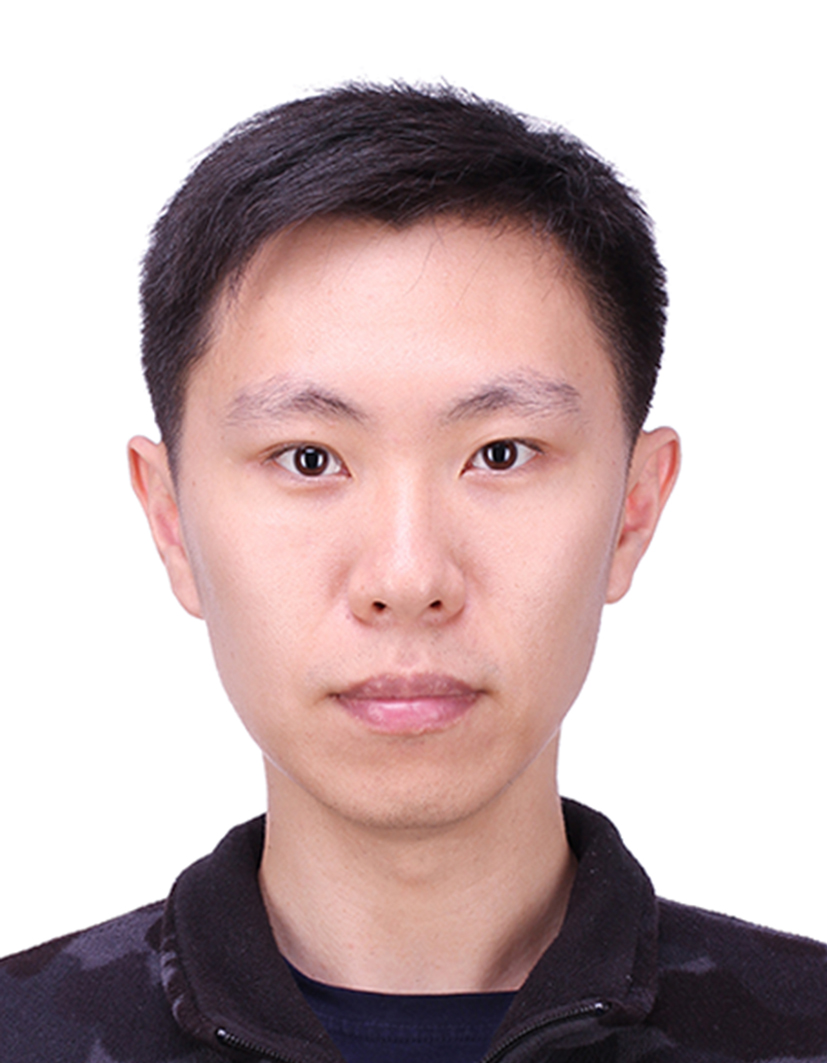}{Mengchen Liu}
{is a Senior Researcher at Microsoft. His research interests include explainable AI and computer vision. He received a B.S. in Electronics Engineering and a Ph.D. in Computer Science from Tsinghua University.
He has served as a PC member and reviewer for various conferences and journals.}

\Author{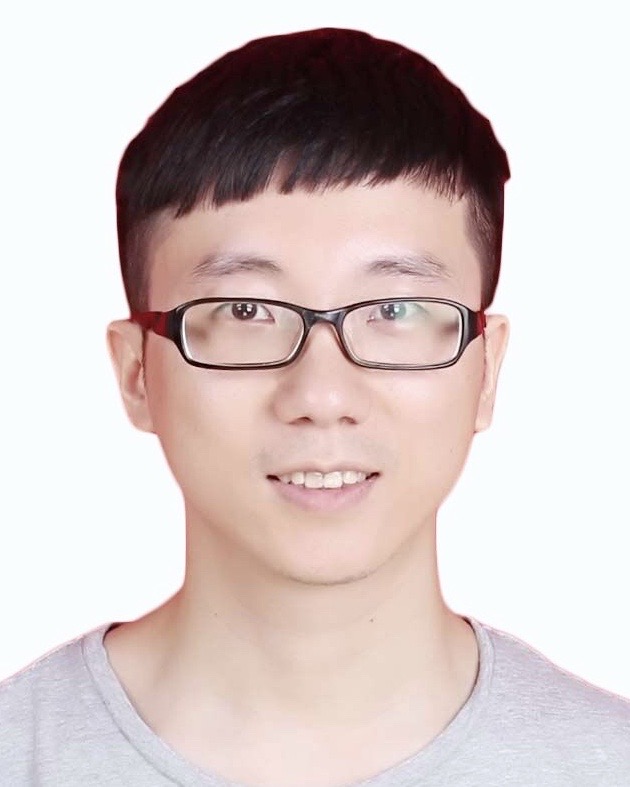}{Jiazhi Xia}
{is an associate professor in the School of Computer Science and Engineering at Central South University. He received his Ph.D. degree in Computer Science from Nanyang Technological University, Singapore in 2011 and obtained his M.S. and B.S. degrees in Computer Science and Technology from Zhejiang University in 2008 and 2005, respectively. His research interests include data visualization, visual analytics, and computer graphics. \\ \\}

\Author{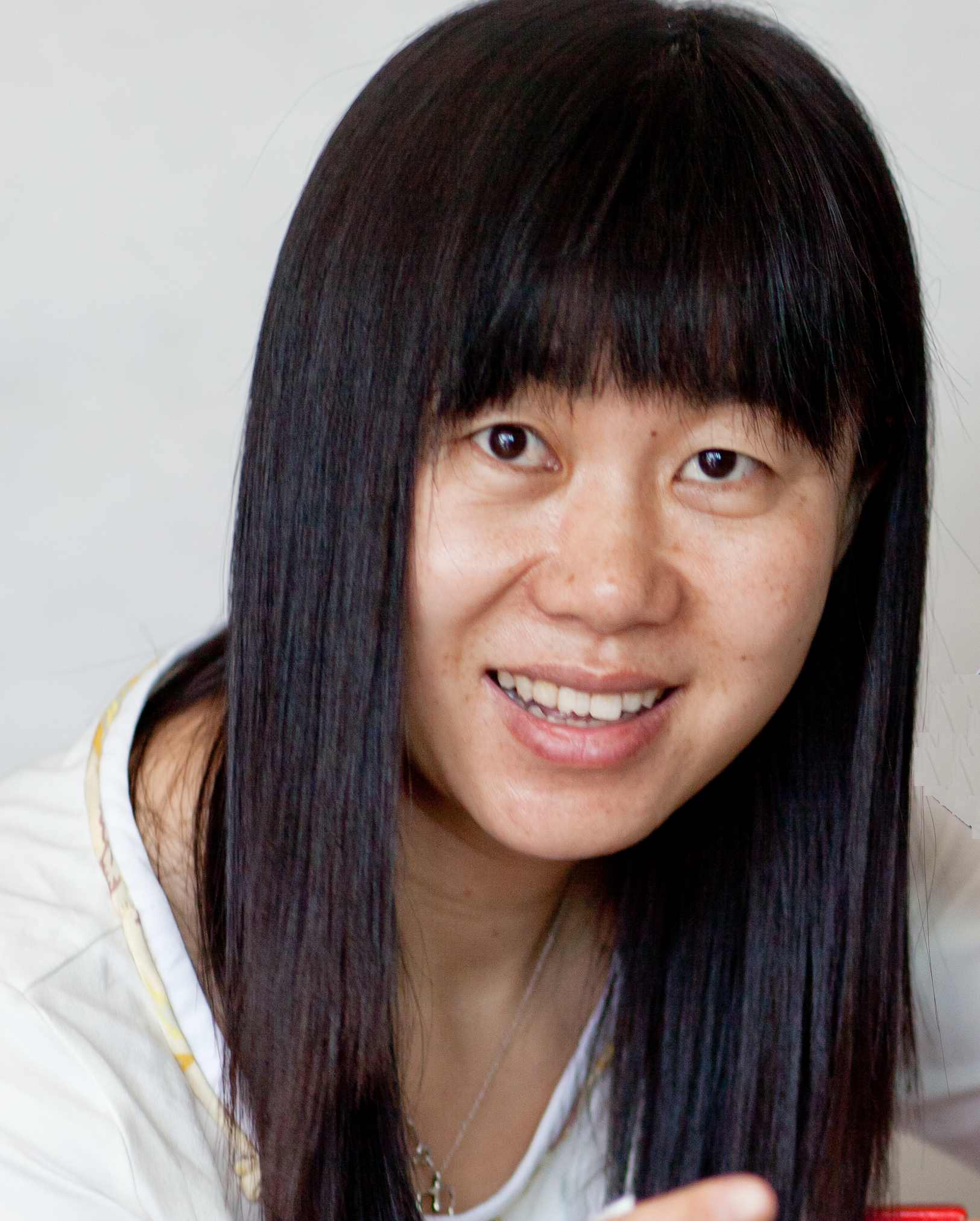}{Shixia Liu}
{is an associate professor at Tsinghua University. Her research interests include visual text analytics, visual social analytics, interactive machine learning, and text mining. She has worked as a research staff member at IBM China Research Lab and a lead researcher at Microsoft Research Asia.
She received a B.S. and M.S. from Harbin Institute of Technology, and a Ph.D. from Tsinghua University.
She is an associate editor-in-chief of IEEE Trans. Vis. Comput. Graph.}
\end{document}

%% file: 0-abstract.tex
\MakeAbstract{
Visual analytics for machine learning has recently evolved as one of the most exciting areas in the \yuanjun{field} of visualization. 
To better identify which research topics are promising and to learn how to apply relevant techniques in visual analytics, we systematically review \mengchen{259} papers published in \yuanjun{the} last \yuanjun{ten} years together with representative works before 2010. 
We build a taxonomy, which includes three first-level categories: techniques before model building, techniques during model building, and techniques after model building. 
Each category is \revision{further characterized by representative analysis tasks, and each task is} exemplified by a set of recent influential works. 
We also discuss and highlight research challenges and promising potential future research opportunities useful for visual analytics researchers.
}

\MakeKeywords{visual analytics; machine learning; data quality; feature selection; model understanding; content analysis}

%% file: 1-introduction.tex
\section{Introduction}\label{sec:introduction}

The recent success of artificial intelligence applications depends on the performance and capabilities of machine learning models~\cite{liu2017towards}. 
In the past ten years, a variety of visual analytics methods have been proposed to \mengchen{make machine learning more explainable, trustworthy, and reliable.}
These research efforts fully combine the advantages of interactive visualization and machine learning techniques to facilitate the analysis and understanding of the major components in the learning process, with an aim to improve performance. 
For example, visual analytics research for explaining the inner workings of deep convolutional neural networks has increased the transparency of deep learning models and has received ongoing, and increasing, \yuanjun{attention} recently~\cite{choo2018visual,hohman2019visual,liu2017towards,Zeiler14VisConvolution}.

The rapid development of visual analytics techniques for machine learning yields an emerging need for a comprehensive review of this area to support the understanding of how visualization techniques are designed and applied to machine learning pipelines.
There have been several initial efforts to summarize the advances in this field \yuanjun{from} different viewpoints.
For example, Liu~\etal~\cite{liu2019bridging} summarized visualization techniques for text analysis. Lu~\etal~\cite{Lu17Predictive17} surveyed \yuanjun{visual analytics techniques} for predictive models.
Recently, Liu~\etal~\cite{liu2017towards} presented a paper on the analysis of machine learning models from the visual analytics viewpoint.
Sacha~\etal~\cite{Sacha19VIS4ML} analyzed a set of example systems and proposed an ontology for visual analytics assisted machine learning.
However, existing surveys either focus on a specific area of machine learning (\eg\ text mining~\cite{liu2019bridging}, predictive models~\cite{Lu17Predictive17}, model understanding~\cite{liu2017towards}), or aim to sketch an ontology~\cite{Sacha19VIS4ML} based on a set of example techniques only.

In this paper, we aim to provide a comprehensive survey of visual analytics techniques for machine learning, which focuses on every phase of the machine learning pipeline. 
\revision{We focus on works in the visualization community. 
Nevertheless, the AI community has also made solid contributions to the study of visually explaining feature detectors in deep learning models.
For example, Selvaraju~\etal~\cite{Selvaraju2017} tried to identify the part of an image to which its classification result is sensitive, by computing class activation maps.
Readers can refer to the surveys of Zhang~\etal~\cite{Zhang2018fitee} and Hohman~\etal~\cite{hohman2019visual} for more details.}
We have collected 259 papers from related top-tier venues in the past ten years through a systematical procedure. Based on the machine learning pipeline, we divide this literature as relevant to three stages: before, during, and after model building. We analyze the functions of visual analytics techniques in the three stages and abstract typical tasks, including improving data quality and feature quality before model building, model understanding, diagnosis, and steering during model building, and data understanding after model building. Each task is illustrated by a set of carefully selected examples. 
We highlight six prominent research directions and open problems in the field of visual analytics for machine learning. We hope that this survey promotes discussion of
\yuanjun{machine learning} related visual analytics techniques and acts as a starting point for practitioners and researchers wishing to develop visual analytics tools for machine learning.

%% file: 2-scope.tex
\section{Survey Landscape}\label{sec:scope}

\subsection{Paper Selection}

In this paper, we focus on visual analytics techniques that \yuanjun{help} to develop explainable, trustworthy, and reliable machine learning applications.
To comprehensively survey visual analytics techniques for machine learning, we performed an exhaustive manual review of relevant top-tier venues in the past ten years (2010-2020): these were InfoVis, VAST, Vis (later SciVis), EuroVis, PacificVis, IEEE TVCG, CGF, and CG\&A.
The manual review was conducted by three Ph.D. candidates with more than two years of research experience in visual analytics.
We followed the manual review process used in a text visualization survey~\cite{liu2019bridging}.
\jiazhi{Specifically, we first considered the titles of papers from these venues to identify candidate papers.}
\jiazhi{Next, we reviewed the abstracts of the candidate papers to further determine whether they concerned visual analytics techniques for machine learning.}
\jiazhi{If the title and abstract did not provide clear information, the full text was gone through to make a final decision.}
In addition to the exhaustive manual review of the above venues, we also searched for the representative related works that appeared earlier or in other venues, such as the Profiler~\cite{kandel2012profiler}.  

After this process, 259 papers were selected. 
Tab.~\ref{tab:category} \jiazhi{presents} detailed statistics.
Due to the increase in machine learning techniques over the past ten years, this field has been attracting ever more research attention.

\begin{figure*}[!tb]
\centering
{\includegraphics[width=\linewidth]{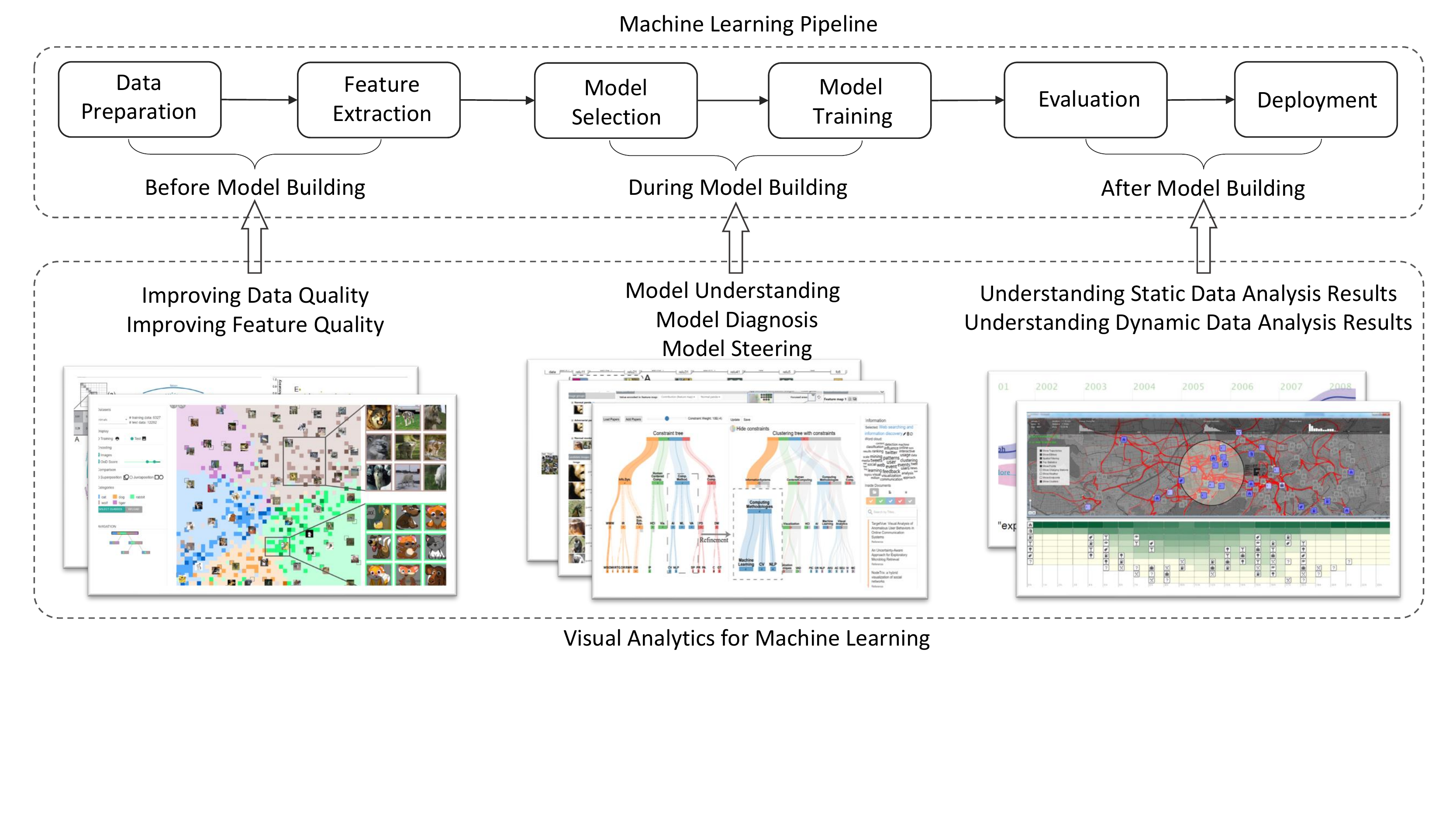}\hfill
\vspace{-2cm}
\caption{An overview of visual analytics research for machine learning.}
\label{fig:system-overview}
}
\end{figure*}

\subsection{Taxonomy}

In this section, we comprehensively analyze the collected visual analytics works to systematically understand the major research trends.
These works are categorized based on a typical machine learning pipeline~\cite{marsland2015machine} used to solve real-world problems.
As shown in Fig.~\ref{fig:system-overview}, such \jiazhi{a} pipeline contains three stages: (1) data pre-processing before model building, (2) machine learning model building, and (3) deployment after the model is built.
Accordingly, visual analytics techniques for machine learning can be mapped into these three stages: techniques before model building, techniques during model building, and techniques after model building.

\subsubsection{Techniques before Model Building}

The major goal of visual analytics techniques before model building is to help model developers better prepare the data for model building.
The quality of the data is mainly determined by the data itself and the features used.
Accordingly, there are two research directions, visual analytics for data quality improvement and feature engineering.


\paragraph{Data quality} can be improved in various ways, such as completing missing data attributes and correcting wrong data labels.
Previously, these tasks were mainly conducted manually or by automatic methods, such as learning-from-crowds algorithms~\cite{hung2015minimizing} which aim to estimate  ground-truth labels from noisy crowd-sourced labels.
To reduce experts' efforts or further improve the results of automatic methods, some works employ visual analytics techniques to interactively improve the data quality.
Tab.~\ref{tab:category} shows that in recent years, this topic has gained increasing research attention.

\paragraph{Feature engineering} is used to select the best features to train the model.
For example, in computer vision, we could use HOG (Histogram of Oriented Gradient) features instead of using raw image pixels.
In visual analytics, interactive feature selection provides an interactive and iterative feature selection process.
In recent years, in the deep learning era, feature selection and construction are mostly conducted via neural networks.  
Echoing this trend, there \yuanjun{is} reducing research attention in recent years (2016--2020) in this direction (see Tab.~\ref{tab:category}).

\subsubsection{Techniques during Model Building}

Model building is a central stage in building a successful machine learning application.
Developing visual analytics methods to \yuanjun{facilitate model building} is also a growing research direction in visualization (see Tab.~\ref{tab:category}).
In this survey, we categorize current methods by their analysis goal: model understanding, diagnosis, and steering.
Model understanding methods aim to visually explain the working mechanisms of a model, such as how changes in parameters influence the model and why the model gives a certain output for a specific input.
Model diagnosis methods target diagnosing errors in model training via interactive exploration of the training process.
Model steering methods are mainly aimed at interactively improving model performance.
For example, to refine a topic model, Utopian~\cite{JaegulChoo2013UTOPIAN} enables users to interactively merge or split topics, and automatically modify other topics accordingly.  

\subsubsection{Techniques after Model Building}

After a machine learning model has been built and deployed, it is crucial to help users (\eg\ domain experts) understand the model output in an intuitive way, to promote trust in the model output. 
To this end, there are many visual analytics methods to explore model output, for a variety of applications.
Unlike methods for model understanding during model building, these methods usually \yuanjun{target} model users rather than model developers.
Thus, the internal workings of a model \yuanjun{are} not illustrated, but the focus is on the intuitive presentation and exploration of model output.
As these methods are often data-driven or application-driven, in this survey, we categorize these methods by the type of data being analyzed, particularly as static data or temporal data.

\begin{table*}[!tb]
\caption{Categories of visual analytics techniques for machine learning and representative works in each category; number of papers given in brackets.}
\centering\label{tab:category}
\small

\begin{tabular}{p{3cm} p{4.4cm} p{6.1cm} p{2cm}}
\toprule  
\multicolumn{2}{l}{\textbf{Technique Category}} & \textbf{Papers} &  \textbf{Trend} \\
\midrule
\multirow{2}{*}{Before Model Building} & Improving Data Quality (31) &
\makecell[l]{\cite{alemzadeh2020visual},~\cite{arbesser2016visplause},~\cite{bauerle2020classifier},~\cite{bernard2019visual},~\cite{bernard2017comparing},~\cite{bernard2018towards},
\cite{bors2019capturing},~\cite{chen2020oodanalyzer},~\cite{dextras2019segmentifier},~\cite{gschwandtner2018know},\\\cite{halter2019vian},~\cite{heimerl2012visual},
\cite{hoferlin2012inter},~\cite{junior2017analytic},~\cite{kandel2012profiler},~\cite{khayat2019vassl},~\cite{kurzhals2016visual},~\cite{lekschas2020peax},\\
\cite{liu2018crowsourcing},~\cite{moehrmann2011improving},~\cite{paiva2014approach},~\cite{park2019cmed},
\cite{park2016c},\cite{rooij2010mediatable},~\cite{snyder2019interactive},~\cite{sperrle2019viana},\\
\cite{stein2016director},
\cite{wang2018graphprotector},~\cite{wang2017utility},~\cite{willett2013identifying},~\cite{xiang2019interactive}} &  \begin{minipage}[b]{\linewidth}\centering\raisebox{-.5\height}{\includegraphics[width=\linewidth]{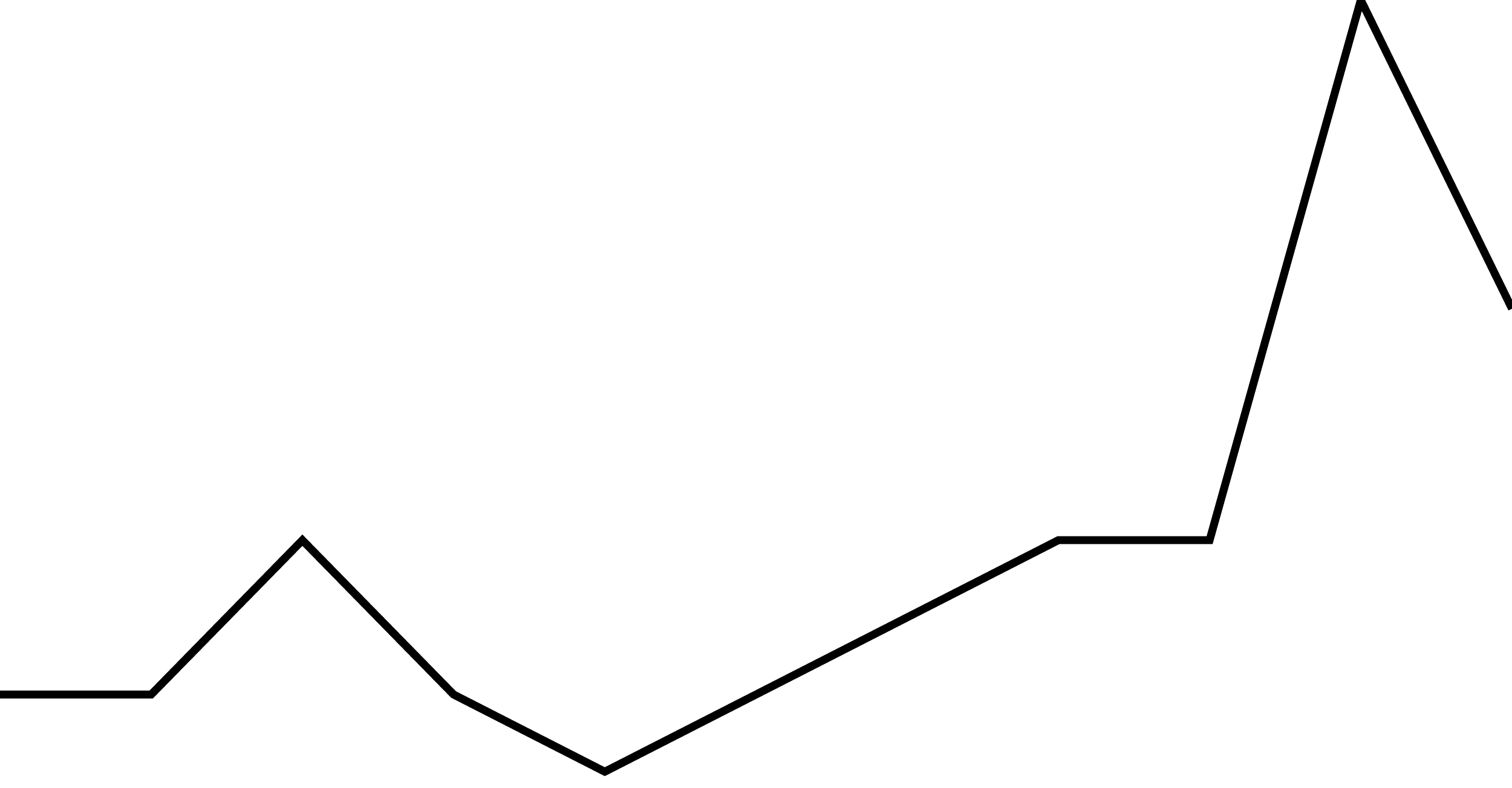}}\end{minipage} \\
\cline {2-4}
& Improving Feature Quality (6) & 
\makecell[l]{\cite{ingram2010dimstiller},~\cite{krause2014infuse},~\cite{may2011guiding},~\cite{muhlbacher2013partition},~\cite{seo2005rank},
\cite{tam2011visualization}} &  \begin{minipage}[b]{\linewidth}\centering\raisebox{-.2\height}{\includegraphics[width=\linewidth]{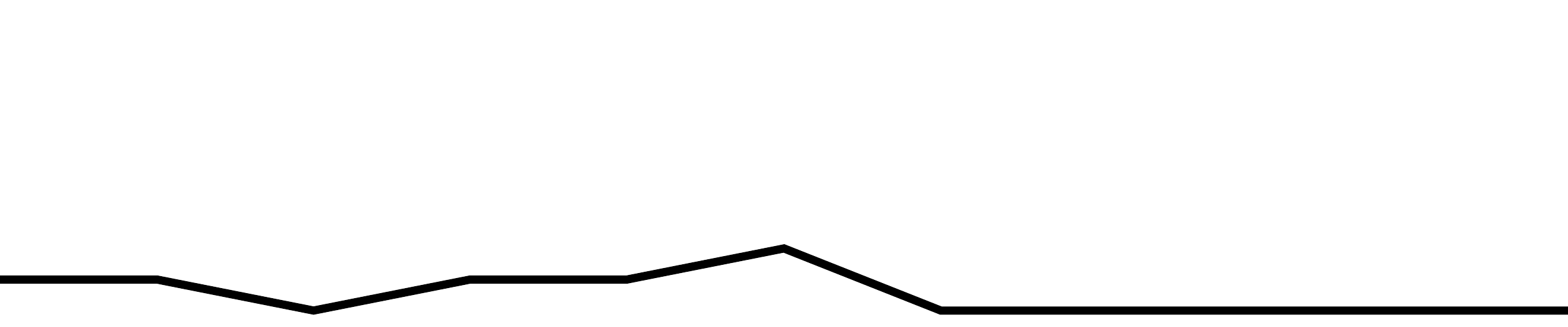}}\end{minipage} \\
\cline {1-4}
\multirow{3}{*}{During Model Building} & Model Understanding (30) & \makecell[l]{\cite{Broeksema2013Decision},~\cite{Cashman2018RNNbow},~\cite{Collaris2020ExplainExplore},~\cite{Eichner2019Making},~\cite{Ferreira2011BirdVis},~\cite{Frhler2016GEMSe},
\cite{Hohman2020Summit},~\cite{Jaunet2020DRLViz},~\cite{Jean2016Dynamic},\\
\cite{Kahng2018ActiVis},~\cite{Kahng2019GANLab},
\cite{Kwon2020DPVis},~\cite{Liu2017TowardsDeepCNN},~\cite{Liu2019NLIZE},~\cite{Migut2011Interactive},~\cite{Ming2017Understanding},
\cite{Ming2019RuleMatrix},\\\cite{Murugesan2019DeepCompareVA},~\cite{Nie2018Visualizing},~\cite{Rauber2017Visualizing},~\cite{Rohlig2015Supporting},
\cite{Scheepens2015Rationale},~\cite{Shen2020Visual},~\cite{Strobelt2018LSTMVis},~\cite{Wang2018GANViz},\\\cite{Wang2019DeepVID},
\cite{Wang2020SCANViz},~\cite{Wongsuphasawat2018Visualizing},~\cite{Zhang2016A},~\cite{Zhao2019iForest}} & \begin{minipage}[b]{\linewidth}\centering\raisebox{-.5\height}{\includegraphics[width=\linewidth]{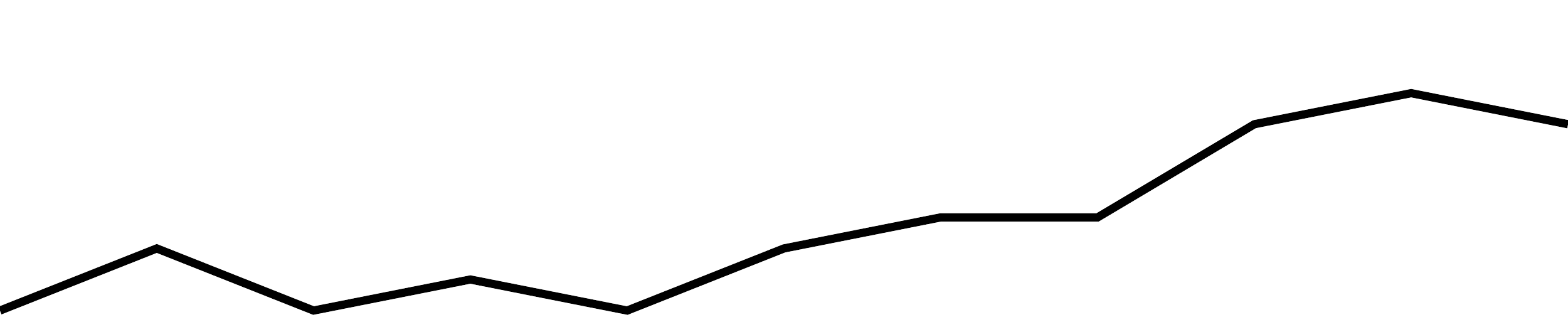}}\end{minipage} \\
\cline {2-4}
& Model Diagnosis (19) & \makecell[l]{\cite{Ahn2020FairSight},~\cite{Alsallakh2014Visual},~\cite{Bilal2018Do},~\cite{Cabrera2019FAIRVIS},~\cite{Cao2020Analyzing},~\cite{Diehl2017Albero},
\cite{Gleicher2020Boxer},~\cite{He2020DynamicsExplorer},~\cite{Krause2017A},~\cite{Liu2018Analyzing},\\
\cite{Liu2018Visual},
\cite{Ma2020Explaining},~\cite{Pezzotti2018DeepEyes},~\cite{Ren2017Squares},~\cite{Spinner2020explAIner},~\cite{Strobelt2019Seq2seq},
\cite{Wang2019DQNViz},~\cite{Wexler2020WhatIf},\\\cite{Zhang2019Manifold}} & \begin{minipage}[b]{\linewidth}\centering\raisebox{-.5\height}{\includegraphics[width=\linewidth]{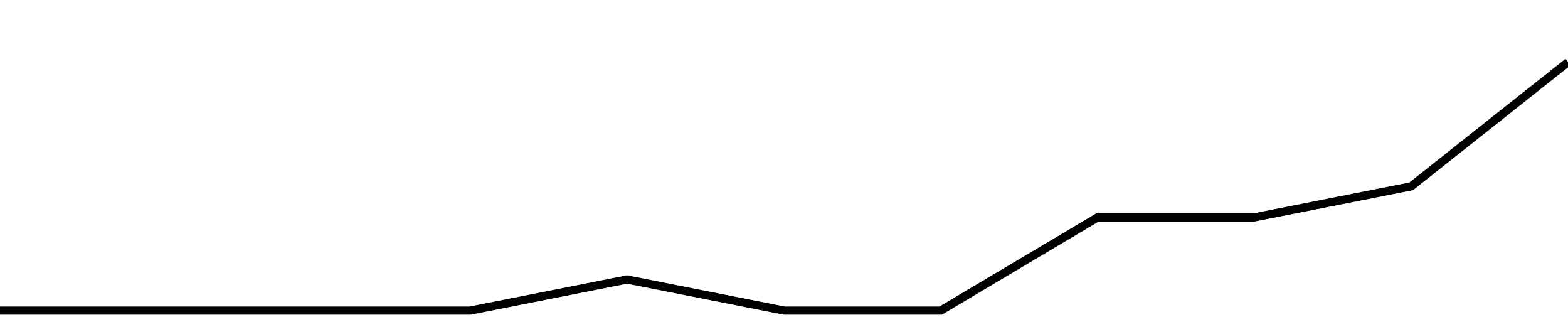}}\end{minipage} \\
\cline {2-4}
& Model Steering (29) & \makecell[l]{
\cite{Bogl2013Visual},~\cite{Cashman2020Ablate},~\cite{Cavallo2018Track},~\cite{Cavallo2019Clustrophile},~\cite{JaegulChoo2013UTOPIAN},~\cite{Das2019BEAMES},
\cite{Dingen2019RegressionExplorer},~\cite{WenwenDou2013HierarchicalTopics},~\cite{ElAssady2020Semantic},\\
\cite{ElAssady2018Progressive},~\cite{ElAssady2019Visual},~\cite{Kim2020ArchiText},
\cite{Kwon2019RetainVis},~\cite{Lee2012iVisClustering},~\cite{Liu2016An},~\cite{Lowe2016Visual},~\cite{MacInnes2010Visual},\\
\cite{Migut2010Visual},~\cite{Ming2020ProtoSteer},~\cite{Muhlbacher2018TreePOD},~\cite{Packer2013Visual},~\cite{Piringer2010HyperMoVal},
\cite{Sacha2018SOMFlow},~\cite{Schultz2013Open},~\cite{vandenElzen2011BaobabView},\\\cite{Vrotsou2019Exploratory},~\cite{Wang2016TopicPanorama},
\cite{Yang2020Interactive},~\cite{Zhao2014LoVis}} & \begin{minipage}[b]{\linewidth}\centering\raisebox{-.5\height}{\includegraphics[width=\linewidth]{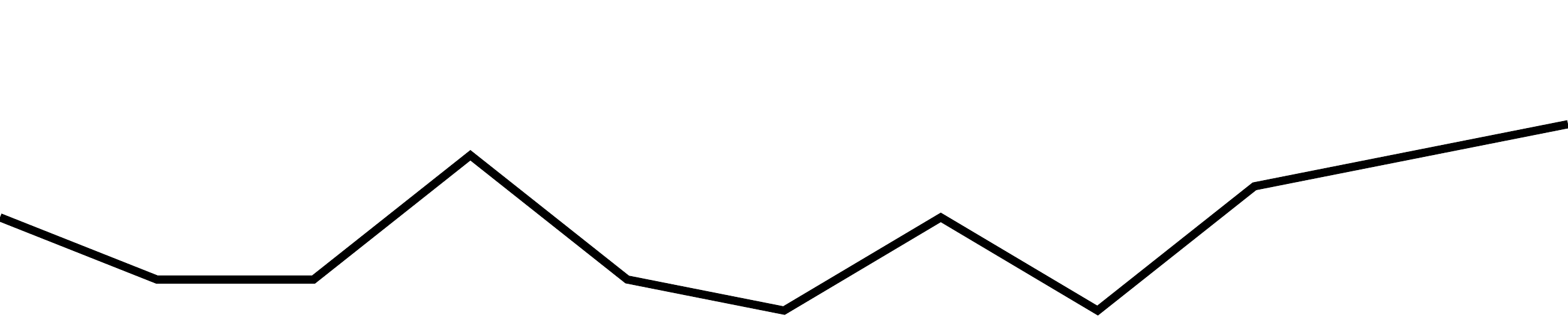}}\end{minipage} \\
\cline {1-4}
\multirow{2}{*}{After Model Building} & Understanding Static Data Analysis Results (43) & \makecell[l]{\cite{alexander2014serendip},~\cite{berger2017cite2vec},~\cite{blumenschein2018smartexplore},~\cite{bradel2014multi},~\cite{broeksema2013visual},~\cite{cao2010facetatlas},
\cite{chandrasegaran2017integrating},~\cite{chen2019lda},~\cite{correll2011exploring},~\cite{dou2015demographicvis},\\\cite{el2016contovi},~\cite{el2018threadreconstructor},
\cite{filipov2019cv3},~\cite{fried2014maps},~\cite{fulda2016timelinecurator},~\cite{glueck2018phenolines},~\cite{gorg2013combining},~\cite{guo2020topic},
\cite{heimerl2016docucompass},\\
\cite{hong2014flda},
\cite{hoque2014convis},~\cite{hu2017visualizing},~\cite{janicke2010soundriver},
\cite{janicke2017interactive},~\cite{jankowska2012relative},~\cite{ji2019visual},
\cite{kakar2019diva},\\
\cite{kim2019topicsifter},
\cite{kim2017topiclens},~\cite{kochtchi2014networks},~\cite{li2018concavecubes},~\cite{liu2015visual},~\cite{liu2014topicpanorama},
\cite{liu2016socialbrands},~\cite{oelke2014comparative},\\
\cite{park2018conceptvector},\cite{paulovich2012semantic},~\cite{shen2018streetvizor},
\cite{von2016comparative},~\cite{wall2018podium},~\cite{xie2019semantic},~\cite{zhang2012hierarchical},~\cite{zhao2012facilitating}}
 & \begin{minipage}[b]{\linewidth}\centering\raisebox{-.5\height}{\includegraphics[width=\linewidth]{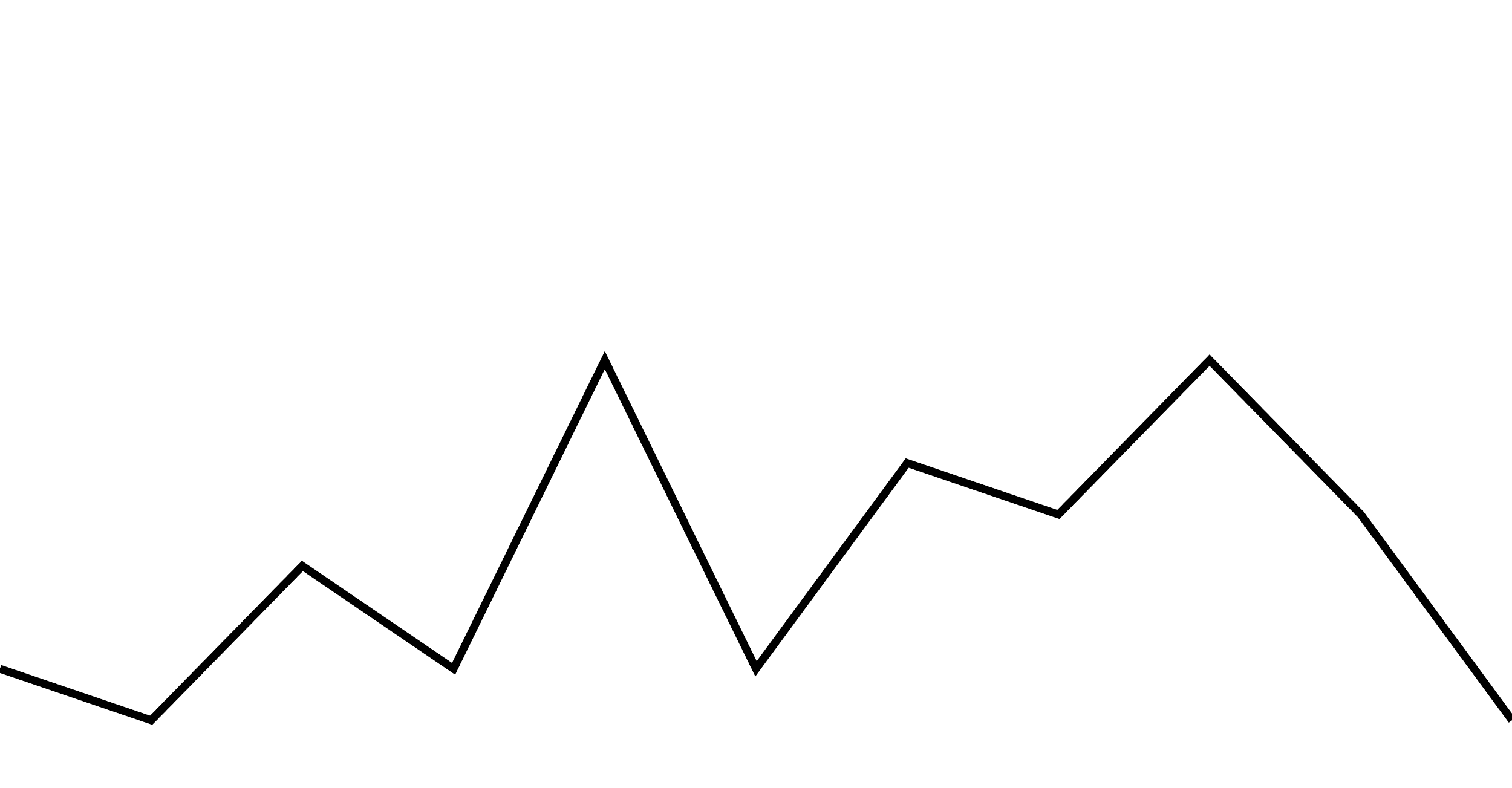}}\end{minipage}  \\
\cline {2-4}
& Understanding Dynamic Data Analysis Results (101) &\makecell[l]{\cite{alsakran2012real},~\cite{alsakran2011streamit},~\cite{Andrienko2019Constructing},~\cite{andrienko2010space},~\cite{andrienko2013scalable},~\cite{blascheck2016visual},~\cite{bogl2017cycle},
\cite{bosch2013scatterblogs2},~\cite{buchmuller2015visual},~\cite{cao2018voila},~\cite{cao2012whisper},\\
\cite{cao2016targetvue},~\cite{chae2012spatiotemporal},
\cite{chen2020viseq},~\cite{chen2017map},~\cite{chen2016d},~\cite{chen2016interactive},~\cite{chen2016dropoutseer},~\cite{chen2018sequence},
\cite{chu2014visualizing},~\cite{cui2011textflow},\\
\cite{cui2014hierarchical},~\cite{di2016allaboard},~\cite{dou2011paralleltopics},~\cite{dou2012leadline},
\cite{du2016eventaction},~\cite{el2017nerex},~\cite{fan2020vista},~\cite{Ferreira2013},~\cite{gobbo2019topic},~\cite{gotz2014decisionflow},\\
\cite{guo2019visual},~\cite{guo2018eventthread},~\cite{gutenko2017anafe},~\cite{havre2002themeriver},~\cite{heimerl2016citerivers},~\cite{itoh2014image},
\cite{itoh2012analysis},~\cite{kamaleswaran2016physioex},~\cite{karduni2017urban},\\
\cite{krueger2019bird},~\cite{krueger2014visual},
\cite{kruger2015semantic},~\cite{lee2019visual},~\cite{leite2018eva},
\cite{li2020semantics},~\cite{li2020weseer},
\cite{li2020galex},\\
\cite{liu2019visual},~\cite{liu2016online},~\cite{liu2012tiara},~\cite{liu2017coreflow},
\cite{liu2017patterns},~\cite{lu2016exploring},~\cite{lu2014business},~\cite{lu2018visual},\\
\cite{luo2012eventriver},
\cite{maciejewski2011forecasting},~\cite{malik2014proactive},~\cite{Miranda2017},~\cite{purwantiningsih2016visual},~\cite{riehmann2019visualizing},
\cite{sacha2017dynamic},~\cite{sarikaya2016visualizing},\\
\cite{shi2014loyaltracker},~\cite{steiger2014visual},~\cite{stopar2019streamstory},
\cite{sultanum2019doccurate},~\cite{sun2014evoriver},
\cite{sung2017exploring},~\cite{thom2012spatiotemporal},~\cite{thom2016can},\\
\cite{Tkachev2019Local},~\cite{vehlow2015visualizing},~\cite{von2016mobilitygraphs},~\cite{wang2012si},~\cite{wang2016ideas},
\cite{wang2018towards},~\cite{wei2010tiara},~\cite{wei2012visual},\\
\cite{wu2020multimodal},~\cite{wu2017mobiseg},
\cite{wu2018streamexplorer},~\cite{wu2014opinionflow},~\cite{wu2016egoslider},~\cite{xie2014vaet},~\cite{xu2017exploring},
\cite{xu2020exploring},\\
\cite{xu2017vidx},~\cite{xu2013visual},~\cite{yu2015iviztrans},~\cite{GarciaZanabria2020crime},
\cite{Zeng2020EmotionCues},~\cite{zeng2020emoco},~\cite{zeng2016visualizing},~\cite{zhang2016visual},\\
\cite{zhang2014visual},
\cite{zhao2014fluxflow},~\cite{zhao2020visual},~\cite{zhou2019visual},~\cite{zhou2017visual}} & \begin{minipage}[b]{\linewidth}\centering\raisebox{-.5\height}{\includegraphics[width=\linewidth]{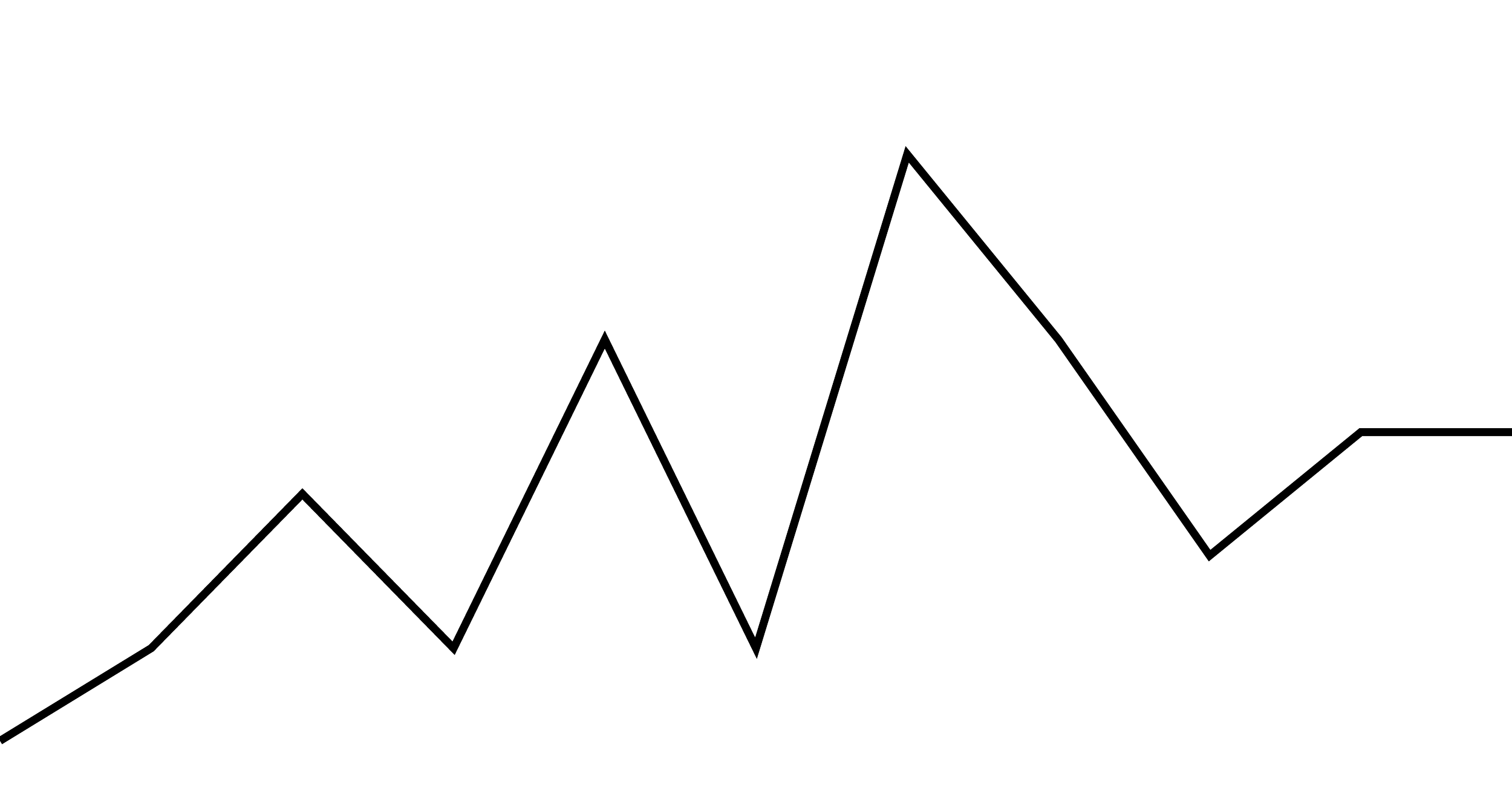}}\end{minipage}  \\
\bottomrule
\end{tabular}
\end{table*}


%% file: 3-before.tex
\section{Techniques before Model Building}\label{sec:before}

Two major tasks required before building a model are data processing and feature engineering. 
\jiazhi{They are critical, as} practical experience indicates that low-quality data and features degrade the performance of machine learning models~\cite{tian2015max,ng2013machine}. 
\jiazhi{Data quality issues include missing values, outliers, and noise in instances and their labels. Feature quality issues include irrelevant features, redundancy between features, etc.}
\jiazhi{While manually addressing these issues is time-consuming, automatic methods may suffer from poor performance.}
Thus, various visual analytics techniques have been developed to reduce experts' efforts 
\jiazhi{as well as to simultaneously improve} the performance of automatic methods of producing high-quality data and features~\cite{Liu2018Steering}.

\subsection{Improving Data Quality}

Data includes instances and their labels~\cite{nilsson1996introduction}.
From this perspective, 
existing efforts for improving data quality either concern instance-level improvement, or label-level improvement.

\begin{figure*}[!tb]
\centering
{\includegraphics[width=\linewidth]{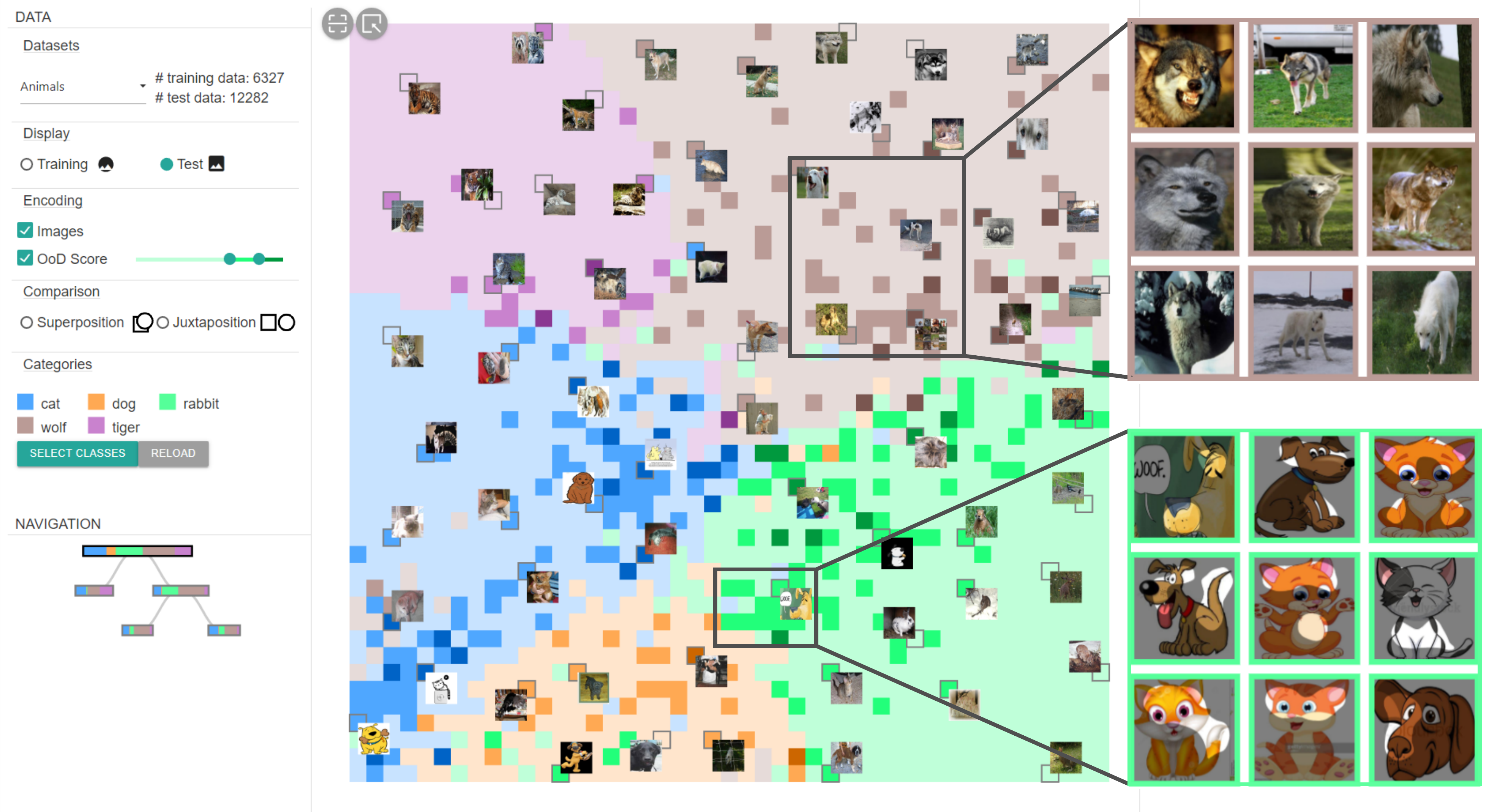}\hfill
\caption{OoDAnalyzer~\cite{chen2020oodanalyzer}, an interactive method to detect out-of-distribution samples and explain them in context.}
\label{fig:oodanalyzer}
}
\end{figure*}

\subsubsection{Instance-level Improvement}
At the instance level, many visual analytics methods focus on detecting and correcting anomalies in data, 
\jiazhi{such as missing values and duplication.}
For example, 
\jiazhi{Kandel~\etal~\cite{kandel2012profiler} proposed Profiler}
to aid the discovery and assessment \jiazhi{of} anomalies in tabular data.
Anomaly detection methods are 
applied to detect data anomalies,
\jiazhi{which are classified into different types subsequently.}
Then, linked summary visualizations are automatically recommended to facilitate the discovery of potential causes and consequences of these anomalies.
VIVID~\cite{alemzadeh2020visual} was developed to handle missing values in longitudinal cohort study data. 
Through multiple coordinated visualizations, experts can identify the root causes of missing values (\eg\ a particular group who do not participate in follow-up examinations), and replace  missing data using an appropriate imputation model.
\mengchen{Anomaly removal is often an iterative process.}
\mengchen{Illustrating} provenance in this iterative process \jiazhi{allows} users to be aware of changes in data quality and to build trust in 
\jiazhi{the} processed data.
Thus, Bors~\etal~\cite{bors2019capturing} \jiazhi{proposed} DQProv Explorer to support the analysis of data processing provenance,
\jiazhi{using a provenance graph to support the navigation of data states and a quality flow to present changes in data quality over time.}
Recently, another type of data anomaly, out-of-distribution (OoD) samples, has received extensive attention~\cite{lakshminarayanan2017simple, lee2017training}.
OoD samples are test samples that are not well covered by training data, which is a major source of model performance degradation.
To tackle this issue, Chen~\etal~\cite{chen2020oodanalyzer} proposed OoDAnalyzer to detect and analyze OoD samples.
An ensemble OoD detection method, combining both high- and low-level features, was proposed to improve detection accuracy.
Based on the detection result, a grid visualization (see Fig.~\ref{fig:oodanalyzer}) is utilized to explore OoD samples in context and explain the underlying reasons for their presence.
\yuanjun{In order to} generate grid layouts at interactive rates during the exploration, a $k$NN-based grid layout algorithm motivated by Hall's theorem was developed.

When considering time-series data, several challenges arise
as time has distinct characteristics that induce specific quality issues that require analysis in a temporal context.
To tackle this issue, Arbesser~\etal~\cite{arbesser2016visplause} proposed a visual analytics system, Visplause, to visually assess time-series data quality.
Anomaly detection results, \eg\ frequencies of anomalies and their temporal distributions, are shown in a tabular layout.
\yuanjun{In order to} address the scalability problem, data are aggregated in a hierarchy based on meta-information, which enables analysis of a group of anomalies (\eg\ abnormal time series of the same type) simultaneously.
\jiazhi{Besides automatically detected anomalies},
KYE~\cite{gschwandtner2018know} also supports the identification of additional anomalies overlooked by automatic methods.
Time-series data 
\jiazhi{are} presented in a heatmap view, where abnormal patterns (\eg\ regions with unusually high values) indicate potential anomalies.
\mengchen{Click stream data are a widely studied kind of time-series data in the field of visual analytics.} 
To better analyze and \mengchen{refine} click stream data, Segmentifier~\cite{dextras2019segmentifier} was proposed to provide an iterative exploration process for segmentation and analysis.
Users can explore segments in three coordinated views at different granularities and refine them by filtering, partitioning, and transformation.
Every refinement step results in new segments, which can be further analyzed and refined. 

\mengchen{To tackle uncertainties in data quality improvement,}
\changjian{Bernard~\etal~\cite{bernard2019visual} developed a visual analytics tool to exhibit the changes in the data and uncertainties caused by different preprocessing methods. 
This tool enables experts to become aware of the effects of these methods and to choose suitable ones, to reduce task-irrelevant parts while preserving task-relevant parts of the data.}


As data have the risk of exposing sensitive information, 
several recent studies have focused on 
preserving data privacy \mengchen{during the data quality improvement process}.
For tabular data, Wang~\etal~\cite{wang2017utility} developed a Privacy Exposure Risk Tree to display privacy exposure risks in the data and a Utility Preservation Degree Matrix to exhibit how \yuanjun{the} utility changes as privacy-preserving operations are applied. 
To preserve privacy in network datasets, Wang~\etal~\cite{wang2018graphprotector} presented a visual analytics system, \emph{GraphProtector}.
To preserve important structures of networks,
node priorities are first specified based on their importance.
Important nodes 
\jiazhi{are} assigned low priorities, reducing the possibility of modifying these nodes.
Based on node priorities and utility metrics, users can apply and compare a set of privacy-preserving operations and choose the most suitable one according to their knowledge and experience.


\subsubsection{Label-level Improvement}
According to whether the data have noisy labels, existing works can be classified as methods either for improving the quality of noisy labels or allowing interactive labeling.

\begin{figure*}[!tb]
\centering
{\includegraphics[width=\linewidth]{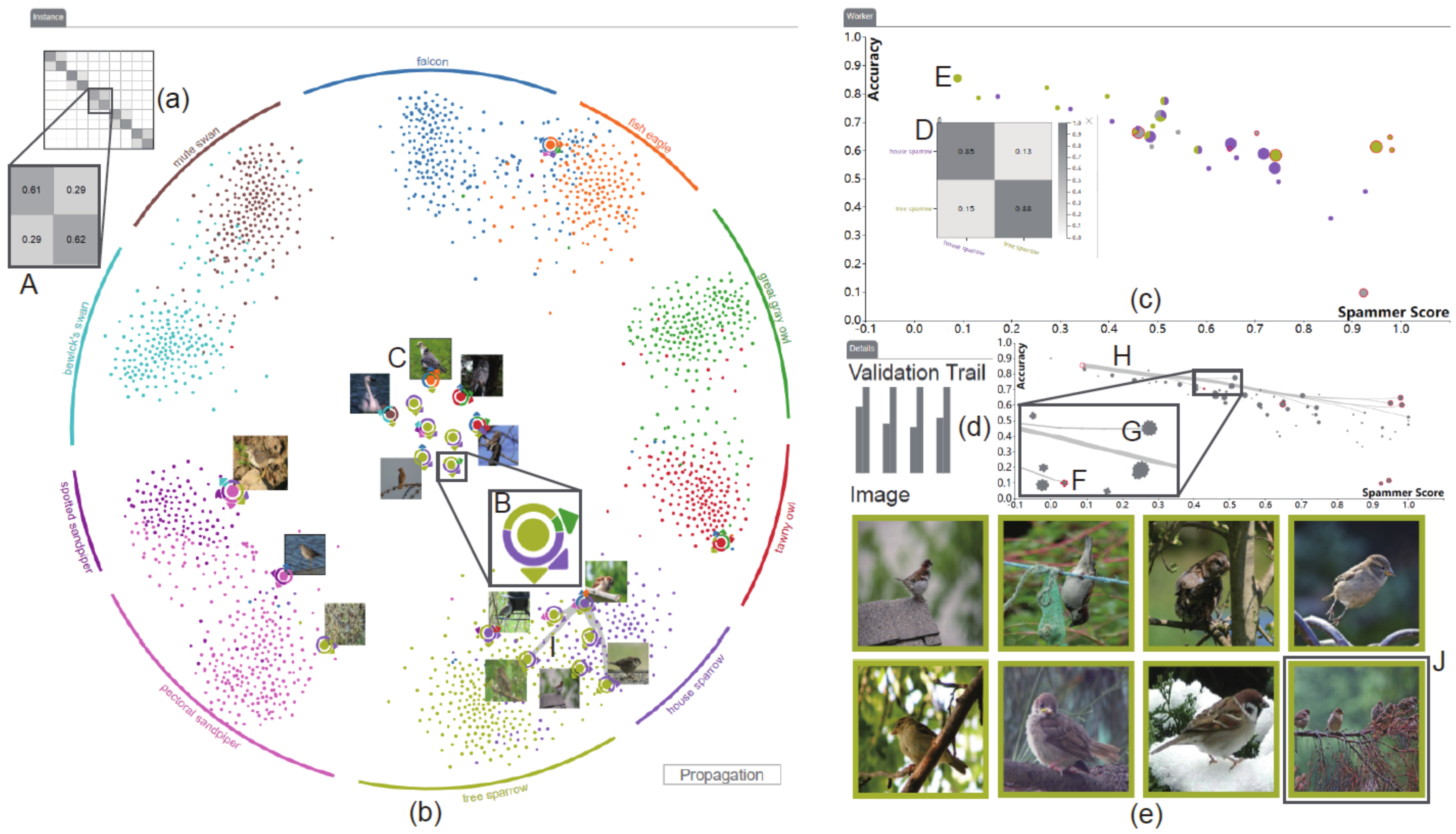}\hfill
\caption{LabelInspect~\cite{liu2018crowsourcing}, an interactive method to verify uncertain instance labels and unreliable workers.}
\label{fig:labelinspect}
}
\end{figure*}

Crowdsourcing provides a cost-effective way to collect labels.
However, annotations provided by crowd workers are usually noisy~\cite{tian2015max, liu2017improving}.
\jiazhi{Many methods have been proposed to remove noise in  labels.}
Willett~\etal~\cite{willett2013identifying} developed a crowd-assisted clustering method to remove redundant explanations provided by crowd workers.
\jiazhi{Explanations are clustered into groups, and the most representative ones are preserved.}
\jiazhi{Park~\etal~\cite{park2016c} proposed C$^2$A that visualizes crowdsourced annotations and worker behavior to help doctors identify malignant tumors in clinical videos.}
\jiazhi{Using C$^2$A,}
doctors can discard most tumor-free video segments and focus on the ones that most likely to contain tumors.
\jiazhi{To analyze the accuracy of crowdsourcing workers, Park~\etal~\cite{park2019cmed} developed CMed that visualizes clinical image annotations by crowdsourcing, and workers' behavior.}
\jiazhi{By clustering workers according to their annotation accuracy and analyzing their logged events, experts are able to find good workers and observe the effects of workers' behavior patterns.}
LabelInspect~\cite{liu2018crowsourcing} was proposed to improve crowdsourced labels by validating uncertain instance labels and unreliable workers. 
Three coordinated visualizations, a confusion (see Fig.~\ref{fig:labelinspect}(a)), an instance (see Fig.~\ref{fig:labelinspect}(b)), and a worker visualization (see Fig.~\ref{fig:labelinspect}(c)), were developed to facilitate the identification and validation of uncertain instance labels and unreliable workers.
Based on expert validation, further instances and workers are recommended for validation by an iterative and progressive verification procedure.

Although the aforementioned methods can effectively improve crowdsourced labels, crowd information is not available in many real-world datasets.
\mengchen{For example, the ImageNet dataset~\cite{russakovsky2015imagenet} only contains the cleaned labels produced by automatic noise removal methods.}
To tackle these datasets, Xiang~\etal~\cite{xiang2019interactive} developed DataDebugger to interactively improve data quality by utilizing user-selected trusted items. 
A hierarchical visualization combined with an incremental projection method and an outlier biased sampling method facilitate the exploration and identification of trusted items.
Based on these identified trusted items,
a data correction algorithm propagates 
\jiazhi{labels}
from \mengchen{trusted items} to the whole dataset.
Paiva~\etal~\cite{paiva2014approach} assumed that instances misclassified by a trained classifier were likely to be mislabeled instances.
Based on this assumption, they employed a Neighbor Joining Tree enhanced by multidimensional projections to help users explore misclassified instances and correct mislabeled ones.
After correction, the classifier is refined using the corrected labels, and a new round of correction starts.
B{\"a}uerle~\etal~\cite{bauerle2020classifier} developed three classifier-guided measures to detect data errors.
Data errors are then presented in a matrix and a scatter plot, allowing experts to reason about and resolve errors.

All the above methods start with a set of labeled data with noise.
However, many datasets do not contain such a label set.
To tackle this issue, many visual analytics methods have been proposed for interactive labeling.
\mengchen{Reducing labeling effort is a major goal of interactive labeling.}
To this end, Moehrmann~\etal~\cite{moehrmann2011improving} used an SOM-based visualization to place similar images together,
 allowing users to label multiple similar images of the same class in one go.
This strategy is also used by Khayat~\etal~\cite{khayat2019vassl} to identify social spambot groups with similar anomalous behavior, 
Kurzhals~\etal~\cite{kurzhals2016visual} to label mobile eye-tracking data,
and Halter~\etal~\cite{halter2019vian} to annotate and analyze primary color strategies used in films.
Apart from placing similar items together, other strategies, like filtering, have also been applied to find items of interest for labeling.
Filtering and sorting are utilized in MediaTable~\cite{rooij2010mediatable} to find similar video segments.
A table visualization is utilized to present video segments and their attributes.
Users can filter out irrelevant segments and sort on attributes to order relevant segments, allowing users  to label several segments of the same class simultaneously.
Stein~\etal~\cite{stein2016director} provided a rule-based filtering engine to find patterns of interest in soccer match videos.
Experts can interactively specify rules through a natural language GUI.

\mengchen{Recently, to enhance the effectiveness \changjian{of} interactive labeling, various visual analytics methods have combined visualization techniques with machine learning techniques, such as active learning.}
The concept of `intra-active labeling' was first introduced by Hoferlin~\etal~\cite{hoferlin2012inter}; it enhances active learning with human knowledge.
Users \yuanjun{are not only able to} query instances \yuanjun{and label them} via active learning, but also \yuanjun{to} understand and steer machine learning models interactively.
\jiazhi{This concept is also used in text document retrieval~\cite{heimerl2012visual}, sequential data retrieval~\cite{lekschas2020peax}, trajectory classification~\cite{junior2017analytic}, identifying relevant tweets~\cite{snyder2019interactive}, and argumentation mining~\cite{sperrle2019viana}.}
For example, to annotate text fragments in argumentation mining tasks, Sperrle~\etal~\cite{sperrle2019viana} 
\jiazhi{developed}
a language model for fragment recommendation.
\jiazhi{A layered visual abstraction is utilized to support five relevant analysis tasks required by text fragment annotation.}
In addition to developing systems for interactive labeling, some empirical experiments were conducted to demonstrate
\jiazhi{their effectiveness.}
\jiazhi{For example, Bernard~\etal~\cite{bernard2017comparing} conducted experiments to show the superiority of user-centered visual interactive labeling over model-centered active learning.
A quantitative analysis~\cite{bernard2018towards} was also performed to evaluate user strategies for selecting samples in the labeling process.}
Results show that in early phases, data-based (\jiazhi{\eg}\ clusters and dense areas) user strategies work well.
However, in later phases, model-based (\eg\ class separation) user strategies  perform better.

\subsection{Improving Feature Quality}
A typical method to improve feature quality is selecting useful features that contribute most to the prediction, \ie\ feature selection~\cite{chandrashekar2014survey}.
A \jiazhi{common} feature selection strategy is to select a subset of features that minimizes the redundancy between 
\jiazhi{them} and maximizes the relevance between them and targets (\eg\ classes of instances)~\cite{may2011guiding}.
Along this line, several methods have been developed to interactively analyze the redundancy and relevance of features.
For example, Seo~\etal~\cite{seo2005rank} proposed a rank-by-feature framework, which ranks features by relevance. 
They visualized ranking results with tables and matrices.
Ingram~\etal~\cite{ingram2010dimstiller} proposed a visual analytics system, DimStiller, which allows users to explore features and their relationships and interactively remove irrelevant and redundant features.
May~\etal~\cite{may2011guiding} proposed SmartStripes to select different feature subsets for different data subsets.
A matrix-based layout is utilized to exhibit the relevance and redundancy of features.
M{\"u}hlbacher~\etal~\cite{muhlbacher2013partition} developed a partition-based visualization for the analysis of the relevance of features or feature pairs.
The features or feature pairs 
\jiazhi{are} partitioned into subdivisions, which allows users to explore the relevance of features (or feature pairs) at different levels of detail.
A parallel coordinates visualization was utilized by Tam~\etal~\cite{tam2011visualization} to identify features that could discriminate between different clusters.
Krause~\etal~\cite{krause2014infuse} ranked features across different feature selection algorithms, cross-validation folds, and classification models.
Users 
are able to interactively select the features and models that lead to the best performance.

\mengchen{Besides selecting existing features, constructing new features is also useful in model building.} 
For example, FeatureInsight~\cite{brooks2015featureinsight} was proposed to construct new features for text classification.
By visually examining classifier errors and summarizing the root causes of these errors, users are able to create new features that can correctly discriminate misclassified documents.
To improve the generalization capability of new features, visual summaries are used to analyze a set of errors instead of individual errors.

%% file: 4-in.tex
\begin{figure*}[!tb]
\centering
{\includegraphics[width=0.9\linewidth]{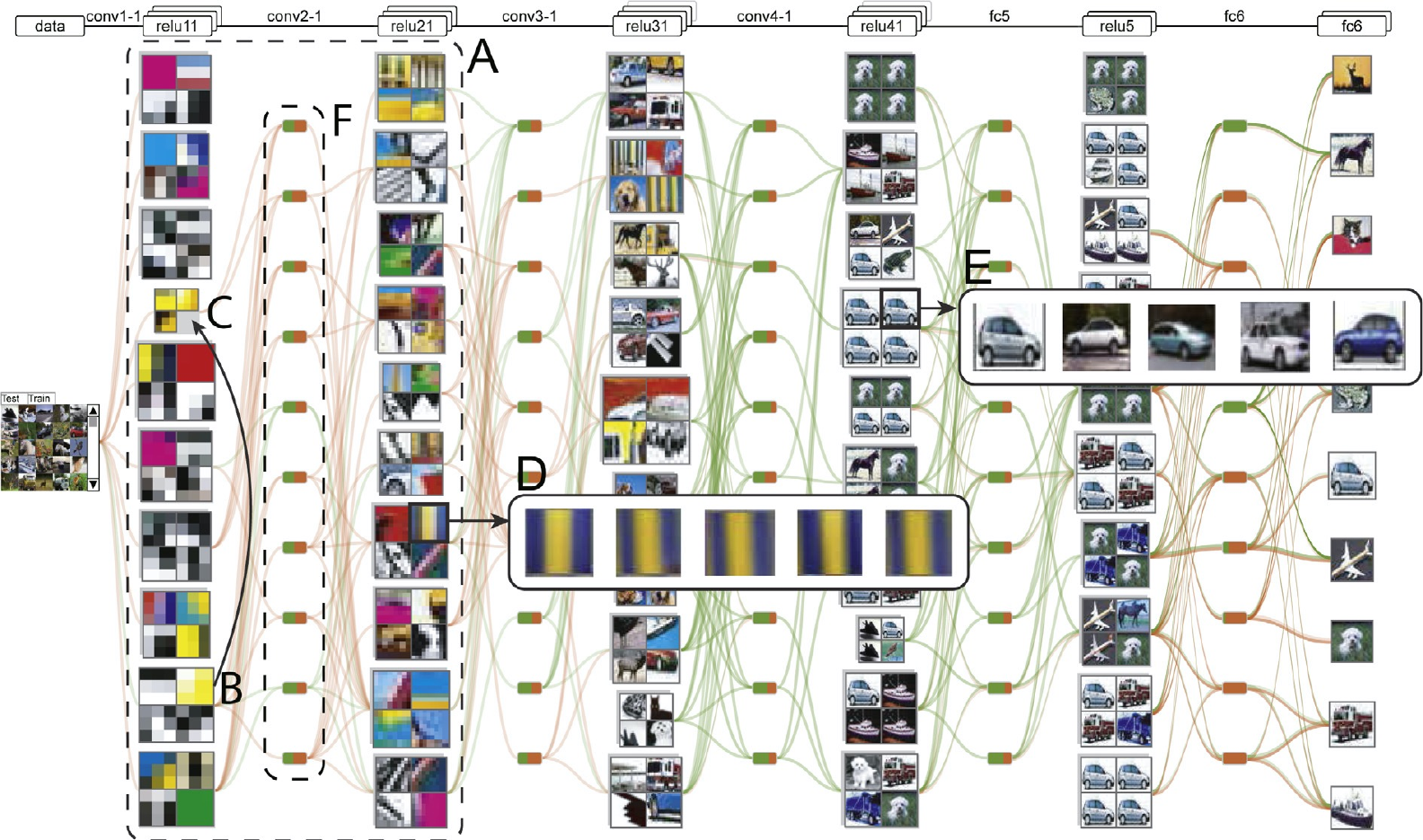}\hfill
\caption{CNNVis~\cite{Liu2017TowardsDeepCNN}, a network-centric visual analytics technique to understand deep convolutional neural networks with millions of neurons and connections.}
\label{fig:CNNVis}
}
\end{figure*}

\section{Techniques during Model Building}


Machine learning models are usually regarded as black boxes because of their lack of interpretability, which hinders their practical use in risky scenarios such as self-driving cars and financial investment.
Current visual analytics techniques in model building explore how to reveal the underlying working mechanisms of machine learning models and then help model developers to build well-performed models.
First of all, model developers require a comprehensive understanding of  models in order to release them from a time-consuming trial-and-error process.
When the training process fails or the model does not provide satisfactory performance, model developers need to diagnose the issues occurring in the training process.
Finally, there is a need to assist in model steering as much time is spent in improving model performance during the model building process.
Echoing these needs, researchers have developed many visual analytics methods to enhance model understanding, diagnosis, and steering~\cite{liu2017towards,choo2018visual}.

\subsection{Model Understanding}
\label{subsec:in_understanding}

Works related to model understanding belong to two classes: those understanding the effects of parameters, and those understanding model behaviours.

\subsubsection{Understanding the Effects of Parameters}
One aspect of model understanding is to inspect how the model outputs change with changes in model parameters.
For example, Ferreira~\etal~\cite{Ferreira2011BirdVis} developed BirdVis to explore the relationships between different parameter configurations and model outputs; these were bird occurrence predictions in their application.
The tool also reveals how these parameters are related to each other in the prediction model.
Zhang~\etal~\cite{zhang2014visual} proposed a visual analytics method to visualize how variables affect  statistical indicators in a logistic regression model.

\subsubsection{Understanding Model Behaviours}
Another aspect is how the model works to produce 
the 
desired outputs. 
There are  three main types of methods used to explain model behaviours, namely network-centric, instance-centric, and hybrid methods.
Network-centric methods aim to explore the model structure
\jiazhi{and} interpret how different parts of the model (\eg\ neurons or layers in convolutional neural networks) cooperate with each other to produce the final outputs.
Earlier 
\jiazhi{works} employ directed graph layouts to visualize the structure of neural networks~\cite{FanYinTzeng2005Opening}, but visual clutter becomes a serious problem as the model structure becoming increasingly complex.
To tackle this problem, Liu~\etal~\cite{Liu2017TowardsDeepCNN} developed CNNVis to visualize deep convolutional neural networks (see Fig.~\ref{fig:CNNVis}).
It  leverages clustering techniques to group neurons with similar roles as well as their connections in order to address visual clutter caused by their huge quantity.
This tool helps experts understand the roles of the neurons and their learned features, and moreover, how low-level features are 
\jiazhi{aggregated} into high-level ones through the network.
Later, Wongsuphasawat~\etal~\cite{Wongsuphasawat2018Visualizing} designed a graph visualization for exploring the machine learning model architecture in Tensorflow~\cite{tensorflow2015-whitepaper}.
They conducted a series of graph transformations to compute a legible interactive graph layout from a given low-level dataflow graph to display the high-level structure of the model.

Instance-centric methods aim to provide instance-level analysis and exploration, as well as understanding the relationships between instances.
Rauber~\etal~\cite{Rauber2017Visualizing} visualized the representations learned from each layer in the neural network by projecting them onto 2D scatterplots.
Users can identify clusters and confusion areas in the representation projections 
\jiazhi{and, therefore, understand} the representation space learned by the network.
Furthermore, they can study how the representation space  evolves during training  so as to understand the network's learning behaviour.
Some visual analytics techniques for understanding recurrent neural networks (RNNs) also adopt such an instance-centric design.
LSTMVis~\cite{Strobelt2018LSTMVis} developed by Strobelt~\etal utilizes parallel coordinates to present the hidden states, to support the analysis of changes in the hidden states over texts.
RNNVis~\cite{Ming2017Understanding} developed by Ming~\etal clusters the hidden state units (each hidden state unit is a dimension of the hidden state vector in an RNN) as memory chips and words as word clouds.
\jiazhi{Their relationships are modeled} as a bipartite graph, which supports sentence-level explanations in RNNs.

Hybrid methods combine the above two methods and leverage both of their strengths.
In particular, instance-level analysis can be enhanced with the context of the network architecture.
Such contexts benefit the understanding of the network's working mechanism.
For instance, Hohman~\etal~\cite{Hohman2020Summit} proposed Summit, to reveal important neurons and critical neuron associations contributing to the model prediction.
It integrates an embedding view to summarize the activations between classes and an attribute graph view to reveal influential connections between neurons.
Kahng~\etal~\cite{Kahng2018ActiVis} proposed ActiVis for large-scale deep neural networks.
It visualizes the model structure with a computational graph and the activation relationships between instances, subsets, and classes using a projected view.

In recent years, there have been some efforts to use a surrogate explainable model to explain model behaviours.
\jiazhi{The major benefit of such methods is that they do not require users to investigate the model itself.}
Thus, they are more useful for those with no or limited machine learning knowledge.
Treating the classifier as a black box, Ming~\etal\cite{Ming2019RuleMatrix} first extracted  rule-based knowledge from the input and output of the classifier.
These rules are then visualized using RuleMatrix, which supports  interactive exploration of the extracted rules by practitioners, improving the interpretability of the model.
Wang~\etal~\cite{Wang2019DeepVID} developed DeepVID to generate a visual interpretation for image classifiers.
Given an image of interest, a deep generative model was first used to generate samples near it.
These generated samples were used to train a simpler and more interpretable model, such as \jiazhi{a} linear regression \jiazhi{classifier},
which helps explain how the original model makes the decision.

\begin{figure*}[!tb]
\centering
{\includegraphics[width=\linewidth]{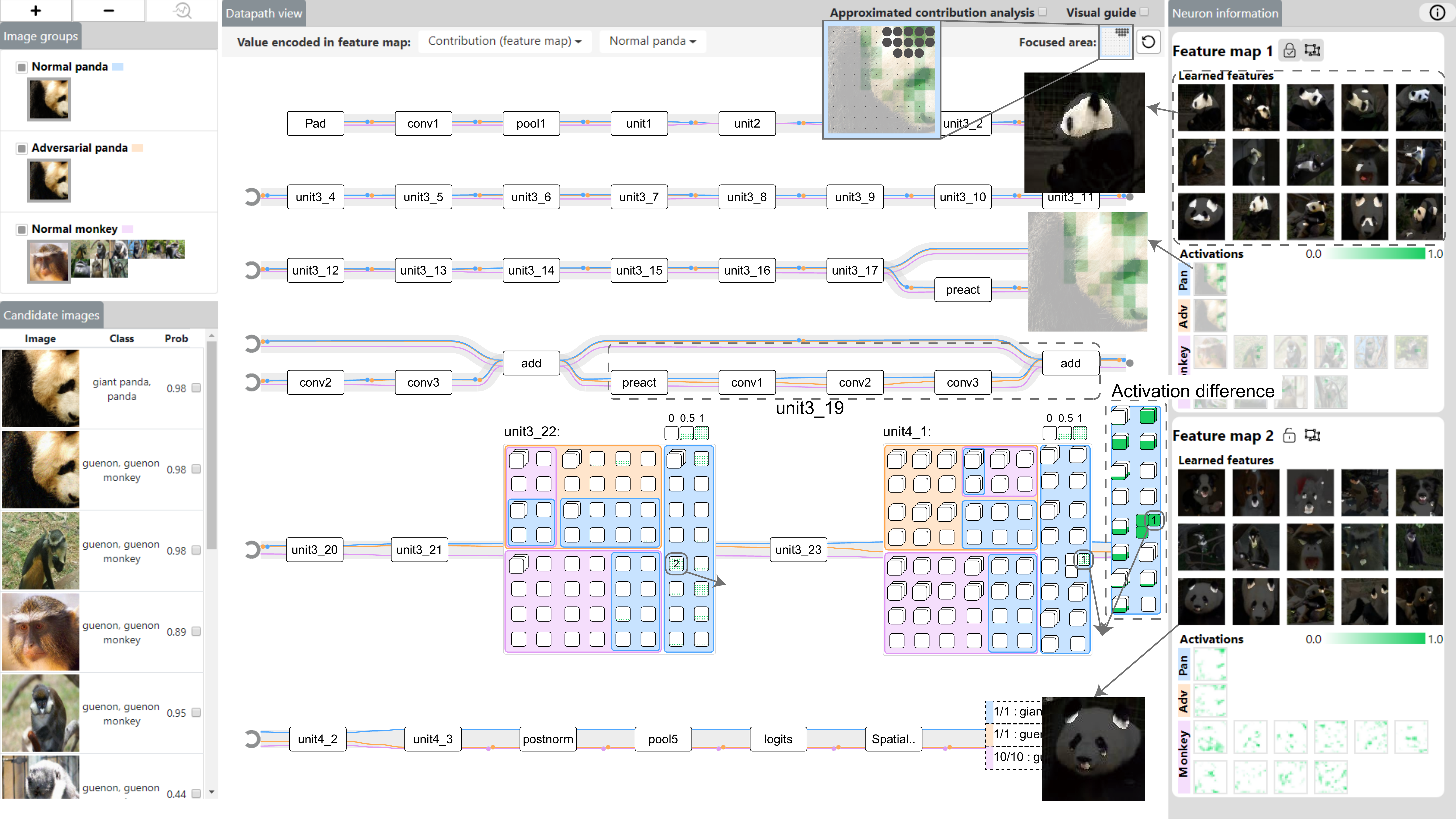}\hfill
\caption{AEVis~\cite{Cao2020Analyzing}, a visual analytics system for analyzing adversarial samples. It shows diverging and merging patterns in the extracted data paths with a river-based visualization, and critical feature maps with a layer-level visualization.}
\label{fig:AEVis}
}
\end{figure*}

\subsection{Model Diagnosis}

Visual analytical techniques for model diagnosis may either analyze the training results or analyze the training dynamics.

\subsubsection{Analyzing Training Results}
Tools have been developed for diagnosing classifiers based on their performance~\cite{Alsallakh2014Visual,Ren2017Squares,Gleicher2020Boxer,Bilal2018Do}.
\jiazhi{For example, Squares~\cite{Ren2017Squares}}
used boxes to represent samples and group them according to their prediction classes.
Using different textures 
\jiazhi{to encode true/false positives/negatives,}
this tool  allows fast and accurate estimation of performance metrics at multiple levels of detail.
Recently, the issue of model fairness has drawn 
growing attention~\cite{Ahn2020FairSight,Cabrera2019FAIRVIS,Wexler2020WhatIf}.
For example, Ahn~\etal~\cite{Ahn2020FairSight} proposed a framework named \textit{FairSight} and implemented a visual analytics system to support the analysis of fairness in ranking problems.
They divided the machine learning pipeline into three phases (data, model, and outcome) and then measured the bias both at  individual  and group levels using different measures.
\jiazhi{Based on these measures,}
developers can iteratively identify 
those features that cause discrimination and remove them from the model.
Researchers are also interested in exploring  potential vulnerabilities 
\jiazhi{in} models that prevent them from being reliably applied to real-world applications~\cite{Ma2020Explaining, Cao2020Analyzing}.
Cao~\etal~\cite{Cao2020Analyzing} proposed AEVis to analyze how adversarial examples fool neural networks.
The system (see Fig.~\ref{fig:AEVis}) takes both normal and adversarial examples as input and extracts their datapaths for model prediction.
It then employs a river-based metaphor to show the diverging and merging patterns of the extracted datapaths, which reveal where the adversarial samples mislead the model.
Ma~\etal~\cite{Ma2020Explaining} designed a series of visual representations from overview to detail to reveal how data poisoning will make a model misclassify a specific sample.
By comparing the distributions of the poisoned and normal training data, experts can deduce the reason for the misclassification of the attacked sample.


\subsubsection{Analyzing Training Dynamics}
Recent efforts have also been concentrated on analyzing the training dynamics.
These techniques are intended for debugging the training process of machine learning models.
For example, DGMTracker~\cite{Liu2018Analyzing} assists experts to discover reasons for the failed training process of deep generative models.
It utilizes a blue-noise polyline sampling algorithm to simultaneously keep the outliers and the major distribution of the training dynamics in order to help experts detect the potential root cause of a failure.
It also employs a credit assignment algorithm to disclose the interactions 
\jiazhi{between} neurons to facilitate the diagnosis of failure propagation.
Attention has also been given to the diagnosis of the training process of deep reinforcement learning.
Wang~\etal~\cite{Wang2019DQNViz} proposed DQNViz for the understanding and diagnosis of deep Q-networks for a Breakout game.
At the overview level, DQNViz presents changes in the overall statistics during the training process with line charts and stacked area charts.
Then at the detail level, it uses segment clustering and a pattern mining algorithm to help experts identify common as well as suspicious patterns in the event-sequences of the agents in Q-networks.
As another example, He~\etal~\cite{He2020DynamicsExplorer} proposed DynamicsExplorer to diagnose an LSTM trained to control a ball-in-maze game.
To support quick identification of where training failures arise, it visualizes ball trajectories with a trajectory variability plot, as well as their clusters using a parallel coordinates plot.

\subsection{Model Steering}

There are two major strategies for model steering: refining the model with human knowledge and selecting the best model from a model ensemble.

\begin{figure*}[!tb]
\centering
{\includegraphics[width=\linewidth]{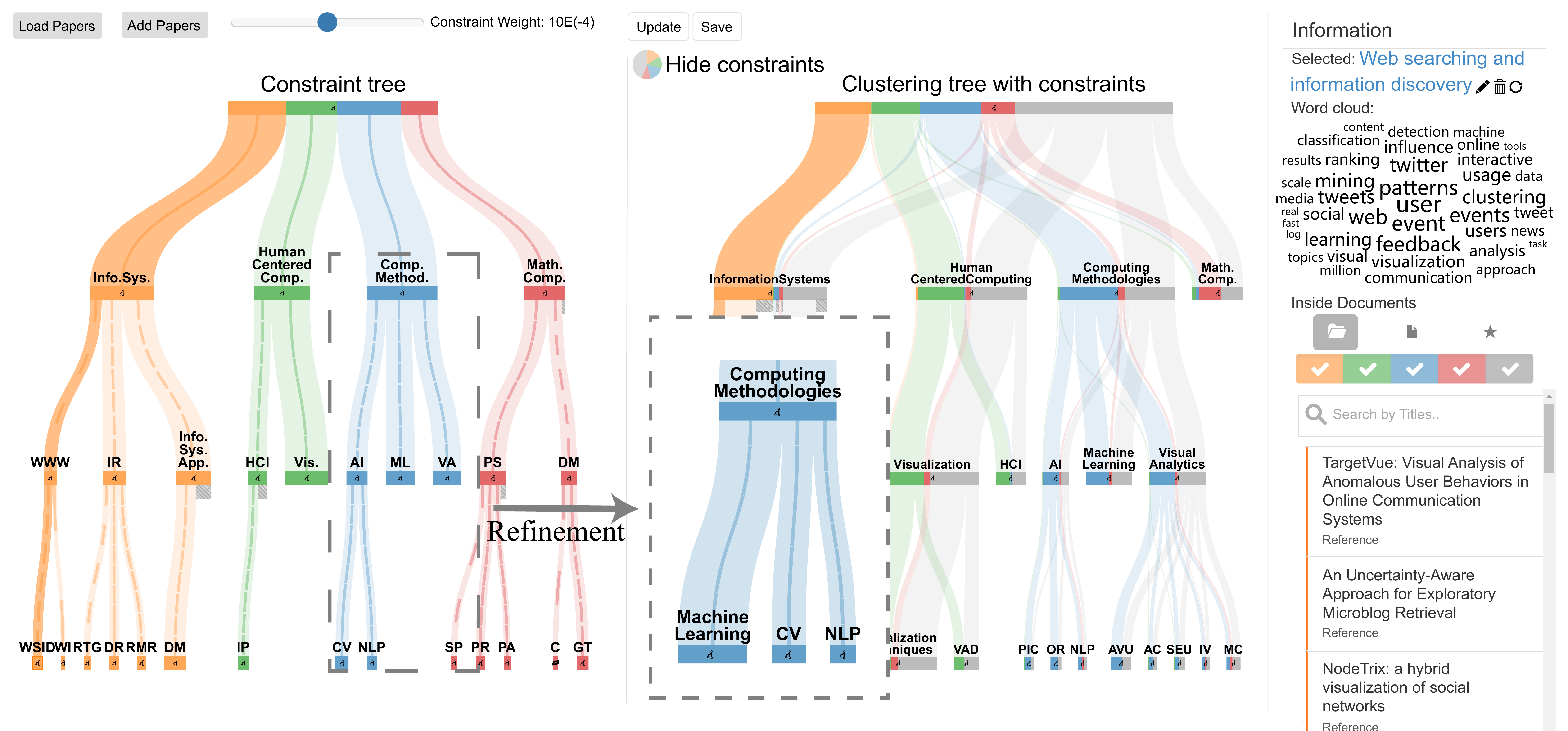}\hfill
\caption{ReVision~\cite{Yang2020Interactive}, a visual analytics system integrating a constrained hierarchical clustering algorithm with an uncertainty-aware, tree-based visualization to help users interactively refine hierarchical topic modeling results.}
\label{fig:ReVision}
}
\end{figure*}

\subsubsection{Model Refinement with Human Knowledge}
Several visual analytics techniques have been developed to place users into the loop of the model refinement process, through flexible interaction.

Users can directly refine the target model with visual analytics techniques.
A typical example is ProtoSteer~\cite{Ming2020ProtoSteer}, a visual analytics system that enables editing prototypes to refine a prototype sequence network named ProSeNet~\cite{Ming2019Interpretable}.
ProtoSteer uses four coordinated views to present the information about the learned prototypes in ProSeNet.
Users can refine these prototypes by adding, deleting, and revising specific prototypes.
\jiazhi{The} model is then retrained with these user-specific prototypes for performance gain.
In addition, van der Elzen~\etal~\cite{vandenElzen2011BaobabView} proposed BaobabView to support experts to construct decision trees iteratively using domain knowledge.
Experts can refine the decision tree with direct operations, including growing, pruning, and optimizing the internal nodes, and can evaluate the refined one with various visual representations.

Besides direct model updates, users can also correct flaws in the results or provide extra knowledge, allowing the model to be updated implicitly to produce improved results based on human feedback.
Several works have focused on incorporating user knowledge into topic models to improve their results~\cite{WenwenDou2013HierarchicalTopics,Wang2016TopicPanorama,JaegulChoo2013UTOPIAN,ElAssady2020Semantic,Yang2020Interactive,Kim2020ArchiText}.
For instance, Yang~\etal~\cite{Yang2020Interactive} presented ReVision that allows users to steer hierarchical clustering results by leveraging an evolutionary Bayesian rose tree clustering algorithm with constraints.
As shown in Fig.~\ref{fig:ReVision}, the constraints and the clustering results are displayed with an uncertainty-aware tree-based visualization to guide the steering of the clustering results.
Users can refine the constraint hierarchy by dragging.
Documents are then re-clustered based on the modified constraints.
Other human-in-the-loop models have also stimulated the development of visual analytic systems 
\jiazhi{to support} such kinds of model refinement.
For instance, Liu~\etal~\cite{Liu2016An} proposed MutualRanker using an uncertainty-based mutual reinforcement graph model to retrieve important blogs, users, and hashtags from microblog data.
It shows ranking results, uncertainty, and its propagation with the help of a composite visualization; users can examine the most uncertain items in the graph and adjust their ranking scores.
The model is incrementally updated by propagating adjustments throughout the graph.

\subsubsection{Model Selection from an Ensemble}
Another strategy for model steering is to select the best model from a model ensemble, which is usually found in clustering~\cite{Schultz2013Open,Packer2013Visual,Cavallo2019Clustrophile} and regression models~\cite{Bogl2013Visual,Lowe2016Visual,Das2019BEAMES,Piringer2010HyperMoVal}.
Clustrophile 2~\cite{Cavallo2019Clustrophile} is a visual analytics system for visual clustering analysis, which guides user selection of appropriate input features and clustering parameters through recommendations based on user-selected results.
BEAMES~\cite{Das2019BEAMES} was designed for multimodel steering and selection in regression tasks.
It creates a collection of regression models by varying algorithms and their corresponding hyperparameters, with further optimization 
\jiazhi{by} interactive weighting of data instances and interactive feature selection and weighting.
Users can inspect them and then select an optimal model according to different aspects of performance, such as their residual scores and mean squared errors.

%% file: 5-after.tex
\section{Techniques after Model Building}

\begin{figure*}[!tb]
\centering
{\includegraphics[width=0.4\linewidth]{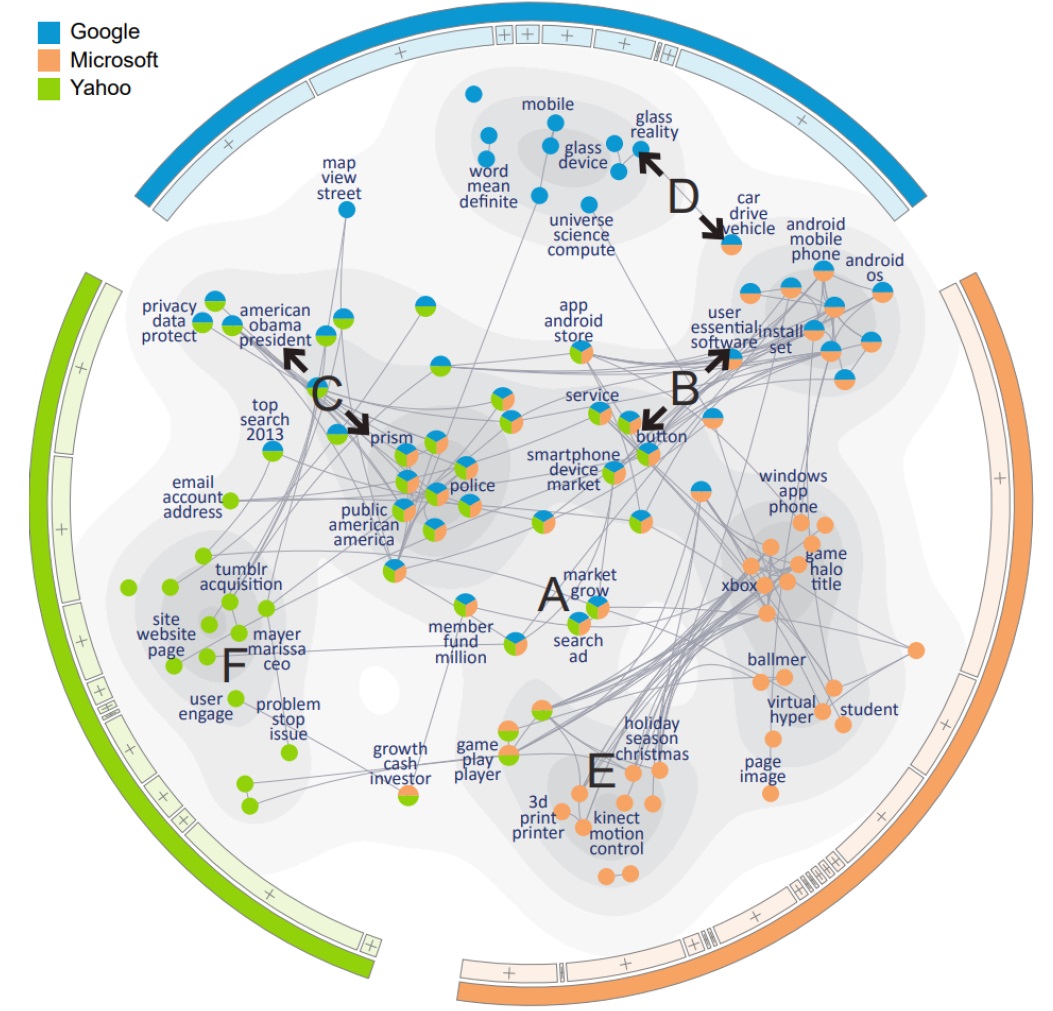}
\hspace{5mm}
\includegraphics[width=0.5\linewidth]{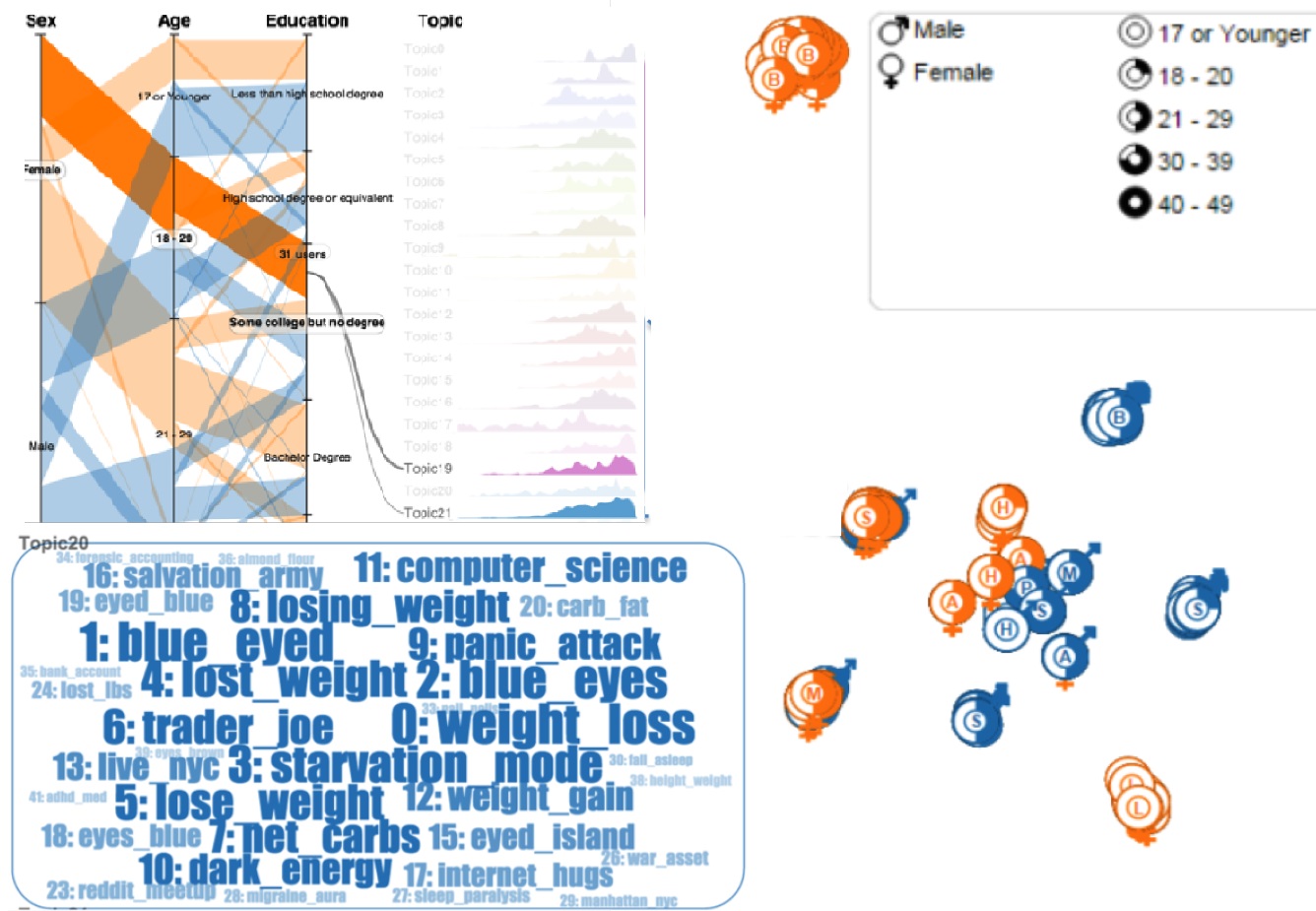}
\hfill
\put(-370,-10){(a)}
\put(-115,-10){(b)}
\caption{Examples of static text visualization.
(a) TopicPanorama~\cite{liu2014topicpanorama} extracts topic graphs from multiple sources and reveals relationships between them using graph layout.
(b) DemographicVis~\cite{dou2015demographicvis} measures similarity between different users after analyzing their posting contents, and reveals their relationships using t-SNE projection.
}
\label{fig:static}
}
\end{figure*}
Existing visual analytics efforts after model building aim to help users understand and gain insights from model outputs, such as high-dimensional data analysis results~\cite{liu2014survey, liu2019bridging}.
As these methods are often data-driven, we categorize the corresponding methods according to the type of data analyzed.
The temporal property of data is critical in visual design.
Thus, we classify methods as those  understanding static data analysis results, and those understanding dynamic data analysis results.
A visual analytics system for understanding static data analysis results usually treats all model output as a large collection and analyzes the static structure.
For dynamic data, in addition to understanding the analysis results at each time point, the system focuses on illustrating the evolution of data over time, which is learned by the analysis model.


\subsection{Understanding Static Data Analysis Results}

We summarize the research on understanding static data analysis according to the type of data. 
Most research focuses on textual data analysis, while fewer works study the understanding of other types of data analysis.

\subsubsection{Textual Data Analysis}
The most widely studied topic is visual text analytics, which tightly integrates interactive visualization techniques with text mining techniques (\eg\ document clustering, topic models, and word embedding) to help users better understand a large amount of textual data~\cite{liu2019bridging}.


Some early works employed simple visualizations to directly convey the results of classical text mining techniques, such as text summarization, categorization, and clustering.
For example, G{\"o}rg~\etal~\cite{gorg2013combining} developed a multi-view visualization consisting of a list view, a cluster view, a word cloud, a grid view, and a document view, to visually illustrate analysis results of document summarization, document clustering, sentiment analysis, entity identification, and recommendation. By combining interactive visualization with text mining techniques, a smooth and informative exploration environment is provided to users.

Most later research has focused on combining well-designed interactive visualization with state-of-the-art text mining techniques, such as topic models and deep learning models, to provide deeper insights into  textual data.
To provide an overview of the relevant topics discussed in multiple sources, Liu~\etal~\cite{liu2014topicpanorama} first utilized a correlated topic model to extract topic graphs from multiple text sources.
A graph matching algorithm is then developed to match the topic graphs from different sources, and a hierarchical clustering method is employed to generate hierarchies of topic graphs.
Both the matched topic graph and hierarchies are fed into a hybrid visualization which consists of a radial icicle plot and a density-based node-link diagram (see Fig.~\ref{fig:static}(a)), to support exploration and analysis of common and distinctive topics discussed in multiple sources.
Dou~\etal~\cite{dou2015demographicvis} introduced DemographicVis to analyze different demographic groups on social media based on the content generated by users.
An advanced topic model, latent Dirichlet allocation (LDA)~\cite{ma2013tag}, is employed to extract topic features from the corpus.
Relationships between the demographic information and extracted features are explored through a parallel sets visualization~\cite{kosara2006parallel}, and different demographic groups are projected onto the two-dimension space based on the similarity of their topics of interest (see Fig.~\ref{fig:static}(b)).
Recently, some deep learning models have also been adopted because of their better performance.
For example, Berger~\etal~\cite{berger2017cite2vec} proposed cite2vec to visualize the latent themes in a document collection via  document usage (\eg\ citations).
It extended a famous word2vec model, the skip-gram model~\cite{mikolov2013distributed}, to generate the embedding for both words and documents by considering the citation information and the textual content together.
The words are projected into a two-dimensional space using t-SNE first, and the documents are projected onto the same space, where both the document-word relationship and document-document relationships are considered simultaneously.

\begin{figure*}[!tb]
\centering
{\includegraphics[width=\linewidth]{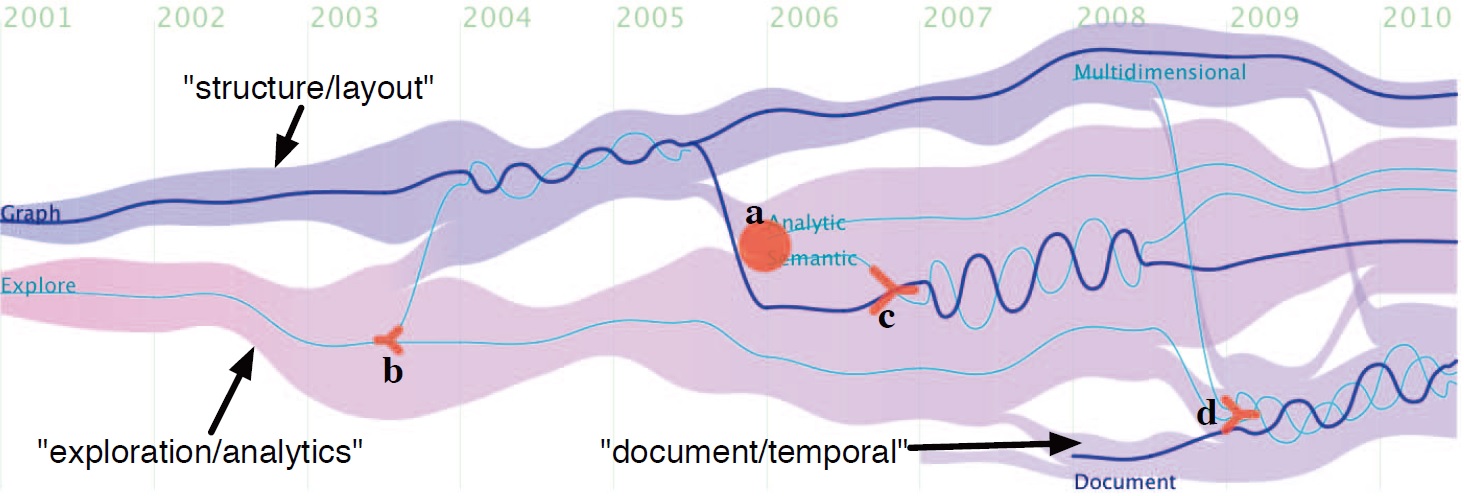}\hfill
\caption{TextFlow~\cite{cui2011textflow} employs a river-based metaphor to show topic birth, death, merging, and splitting.}
\label{fig:textflow}
}
\vspace{-3mm}
\end{figure*}

\subsubsection{Other Data Analysis}
 \mengchen{In addition to textual data, other types of data have also been studied.}
For example, Hong~\etal~\cite{hong2014flda} analyzed flow fields through an LDA model by defining pathlines as documents and features as words, respectively.
After modeling, the original pathlines and  extracted topics were projected into a two-dimensional space using multidimensional scaling, and several previews were generated to render the pathlines for important topics.
Recently, a visual analytics tool, SMARTexplore~\cite{blumenschein2018smartexplore}, was developed to help analysts find and understand interesting patterns within and between dimensions, including correlations, clusters, and outliers.
To this end, it tightly couples a table-based visualization with pattern matching and subspace analysis.

\subsection{Understanding Dynamic Data Analysis Results}

In addition to understanding the results of static data analysis, it is also important to investigate and analyze how latent themes in data change over time.
For example, a system can help politicians to make timely decisions if it provides an overview of major public opinions on social media and how they change over time.
Most existing works focus on understanding the analysis results of a data corpus where each data item is associated with a time stamp. According to whether the system supports the analysis of streaming data, we may further classify existing works on visual dynamic data analysis as offline and online.
\mengchen{In offline analysis, all data are available before analysis, while online analysis tackles streaming data that is incoming during the analysis process.}

\subsubsection{Offline Analysis} 
Offline analysis research can be classified according to the analysis task: topic analysis, event analysis, and trajectory analysis.

Understanding topic evolution in a large text corpus over time is an important topic, attracting much attention.
Most existing works adopt a river metaphor to convey changes in the text corpus over time.
ThemeRiver~\cite{havre2002themeriver} is one of the pioneering works, using the river metaphor to reveal changes in the volumes of different themes.
To better understand the content change of a document corpus, TIARA~\cite{liu2012tiara,wei2010tiara} utilizes an LDA model~\cite{blei2003latent} to extract  topics from the corpus and reveal their changes over time.
However, only observing volumes and content change is not enough for  complex analysis tasks where users want to explore relationships between different topics and their changes over time.
Therefore, later works have focused on understanding  relationships  between topics (\eg\ topic splitting and merging) and their evolving patterns over time.
For example, 
Cui~\etal~\cite{cui2011textflow} first extracted  topic splitting and merging patterns from a document collection using an incremental hierarchical Dirichlet process model~\cite{Yee2006hierarchical}. Then a river metaphor with a set of well-designed glyphs was developed to visually illustrate the aforementioned topic relationships and their dynamic changes over time.
Xu~\etal~\cite{xu2013visual} leveraged a topic competition model to extract  dynamic competition between topics and the effects of opinion leaders on social media.
Sun~\etal~\cite{sun2014evoriver} extended the competition model to a `coopetition' (cooperation and competition) model to help understand the more complex interactions between  evolving topics.
Wang~\etal~\cite{wang2016ideas} proposed IdeaFlow, a visual analytics system for learning the lead-lag relationships across different social groups over time.
However, these works use a flat structure to model  topics, which hampers their usage in the era of big data for handling large-scale text corpora.
Fortunately, there are already initial efforts in coupling hierarchical topic models with interactive visualization to favor the understanding \yuanjun{of} the main content in a large text corpus.
For example, Cui~\etal~\cite{cui2014hierarchical} extract a sequence of topic trees using an evolutionary Bayesian rose tree algorithm~\cite{wang2013mining} and then calculates the tree cut for each tree.
These tree cuts are used to approximate the topic trees and display them in a river metaphor, which also reveals  dynamic relationships between the topics, including topic birth, death, splitting, and merging.

Event analysis targets revealing common or semantically important sequential patterns in  ordered sequences of events~\cite{guo2018eventthread,janicke2010soundriver,liu2017patterns,luo2012eventriver}.
To facilitate visual exploration of large scale event sequences and pattern discovery, several visual analytics methods have been proposed.
For example, Liu~\etal~\cite{liu2017patterns} developed a visual analytics method for click stream data.
Maximal sequential patterns are discovered and pruned from the click stream data.
The extracted patterns and original data are well illustrated at four granularities: patterns, segments, sequences, and events. 
Guo~\etal~\cite{guo2018eventthread} developed EventThread, which uses a tensor-based model to transform the event sequence data into an $n$-dimensional tensor.
Latent patterns (threads) are extracted with a tensor decomposition technique, segmented into stages, and then clustered.
These threads are represented as segmented linear stripes, and a line map metaphor is used to reveal the changes between different stages.
Later, EventThread was extended to overcome the limitation of the fixed length of each stage~\cite{guo2019visual}.
The authors proposed an unsupervised stage analysis algorithm to effectively identify the latent stages in event sequences. 
Based on this algorithm, an interactive visualization tool was developed to reveal and analyze the evolution patterns across stages.

\begin{figure*}[!tb]
\centering
{\includegraphics[width=\linewidth]{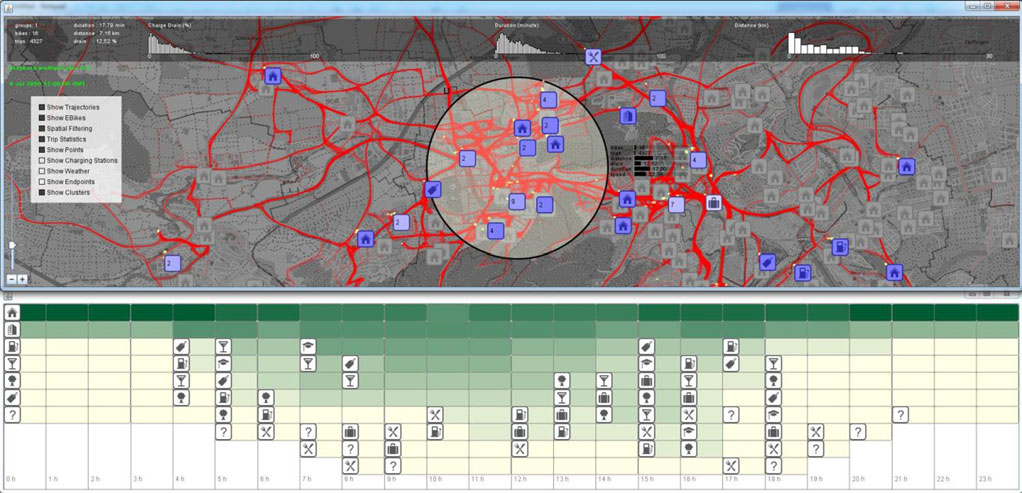}\hfill
\caption{Kruger~\etal~\cite{kruger2015semantic} enrich trajectory data semantically.
Frequent routes and destinations are visualized in the geographic view (top), while the frequent temporal patterns are mined and displayed in the temporal view (bottom).}
\label{fig:trajectory}
}
\end{figure*}

Other works focus on understanding movement data (\eg\ GPS records) analysis results.
Andrienko~\etal~\cite{andrienko2013scalable} extracted  movement events from trajectories and then performed spatio-temporal clustering for aggregation.
These clusters are visualized using spatio-temporal envelopes to help analysts find  potential traffic jams in the city.
Chu~\etal~\cite{chu2014visualizing} adopted an LDA model for mining  latent movement patterns in taxi trajectories.
The movement of each taxi, represented by the traversed street names, was regarded as a document.
Parallel coordinates were used to visualize the distribution of streets over topics, where each axis represents a topic, and each polyline represents a street.
The evolution of the topics was visualized as topic routes that connect similar topics between adjacent time windows.
More recently, Zhou~\etal~\cite{zhou2019visual} treated origin-destination flows as words and trajectories as paragraphs, respectively.
Therefore, a word2Vec model was used to generate the vectorized representation for each origin-destination flow.
t-SNE was then employed to project the embedding of the flows into two-dimensional space, where analysts can check the distributions of the origin-destination flows and select some for display on the map.
Besides directly analyzing the original trajectory data, other papers try to augment the trajectories with auxiliary information to reduce the burden on visual exploration.
Kruger~\etal~\cite{kruger2015semantic} clustered  destinations with DBScan and then used Foursquare to provide detailed information about the destinations (\eg\ shops, university, residence).
Based on the enriched data,  frequent patterns were extracted and displayed in the visualization (see Fig.~\ref{fig:trajectory});  icons on the time axis help understand these patterns.
Chen~\etal~\cite{chen2016interactive} mined  trajectories from  geo-tagged social media and displayed keywords extracted from the text content,  helping users explore the semantics of trajectories.


\subsubsection{Online Analysis}
Online analysis is especially necessary for streaming data, such as text streams.
As a pioneering work for online analysis of text streams, Thom~\etal~\cite{thom2012spatiotemporal} proposed ScatterBlog to analyze geo-located tweet streams.
The system uses Twitter4J to get streaming tweets and extracts location, time, user ID, and tokenized terms in the tweets.
To efficiently analyze a tweet stream, an incremental clustering algorithm was employed to cluster similar tweets.
Based on the clustering results, spatio-temporal anomalies were detected and reported to  users in real-time.
To reduce user effort for filtering and monitoring in ScatterBlogs, Bosch~\etal~\cite{bosch2013scatterblogs2} proposed ScatterBlogs2, which enhanced ScatterBlogs with machine learning techniques.
In \yuanjun{particular}, an SVM-based classifier was built for filtering tweets of interest, and an LDA model was employed to generate a topic overview.
To efficiently handle high-volume text streams, Liu~\etal~\cite{liu2016online} developed TopicStream to help users analyze hierarchical topic evolution in high-volume text streams.
In TopicStream, an evolutionary topic tree was built from text streams, and a tree cut algorithm was developed to reduce visual clutter and enable users to focus  on topics of interest. 
Combining a river metaphor and a visual sedimentation metaphor, the tool effectively illustrates the overall hierarchical topic evolution as well as how newly arriving textual documents are gradually aggregated into the existing topics over time. 
Triggered by TopicStream, Wu~\etal~\cite{wu2018streamexplorer} developed StreamExplorer, which enables the tracking and comparison of a social stream.
In particular, an entropy-based event detection method was developed to detect  events in the social media stream.
They are further visualized in a multi-level visualization, including a glyph-based timeline, a map visualization, and interactive lenses.
In addition to text streams, other types of streaming data are also analyzed.
For example, Lee~\etal~\cite{lee2019visual} employed a long short-term memory model for road traffic congestion forecasting and visualized the results with a Volume-Speed Rivers visualization.
Propagation of  congestion \yuanjun{was} also extracted and visualized,  helping analysts understand causality within the detected congestion.

%% file: 6-opportunity.tex
\section{Research Opportunities}
Although visual analytics research for machine learning \yuanjun{has} achieved promising results in both academia and real-world applications, there are still several long-term research challenges. 
Here, we discuss and highlight major challenges and potential research opportunities in this area.

\subsection{Opportunities before Model Building}
\subsubsection{Improving Data Quality for Weakly Supervised Learning}
Weakly supervised learning builds 
models from data with quality issues, 
\jiazhi{including inaccurate labels, incomplete labels, and inexact labels.}
Improving data quality can boost the performance \jiazhi{of} weakly supervised learning models~\cite{li2020towards}.
Most existing methods focus on inaccurate data (\eg\ noisy crowdsourced annotations and label errors) quality issues, and interactive labeling related to incomplete data (\eg\ none or only a few data are labeled) quality issues.
However, \yuanjun{fewer} efforts are devoted to the better exploitation of unlabeled data related to incomplete data quality issues as well as inexact data (\eg\ coarse-grained labels that are not exact as required) quality issues. 
This paves the way for potential future research. 

\jiazhi{Firstly, }the potential for visual analytics techniques to address the incompleteness issue is not fully exploited.
For example, improving the quality of unlabeled data is critical for semi-supervised learning~\cite{li2020towards, li2016graph}, 
which is tightly combined with a small amount of labeled data during training to infer the correct mapping from the data set to the label set. 
One typical example is graph-based semi-supervised learning~\cite{li2016graph}, which depends on the relationship between labeled and unlabeled data.
Automatically constructed relationships (graphs) are sometimes poor in quality, resulting in model performance degradation.
A major cause behind these poor-quality graphs is that automatic graph construction methods usually rely on global parameters (\eg\ a global $k$ value in the $k$NN graph construction method), which may  be locally inappropriate.
As a consequence, it is necessary to utilize visualization to illustrate how labels are propagated along graph edges, to facilitate  understanding of how local graph structures affect model performance.
Based on such understanding, experts can adaptively modify the graph to gradually create a higher-quality graph.

Secondly, although the inexact data quality issue is common in real-world applications~\cite{zhou2018brief}, it has received little attention from the field of visual analytics.
This issue refers to the situation where labels are inexact, \eg\ coarse-grained labels, such as arise in computed tomography (CT) scans.
The labels of CT scans usually come from corresponding diagnosis reports that describe whether patients have certain medical problems (\eg\ a tumor).
For a CT scan with tumors, we only know that one or more slices in the scan contain tumors. 
However, we do not know which slices contain tumors as well as the exact tumor locations in these slices.
Although various machine learning methods~\cite{foulds2010review, zhou2006multi} 
have been proposed to learn from such coarse-grained labels, they may lead to poor performance~\cite{li2020towards} due to the lack of exact information. 
\jiazhi{Fine-grained validation is still required to improve data quality.}
To this end, one potential solution is to combine interactive visualization with learning algorithms to better illustrate the root cause of bad performance by examining the overall data distribution and the wrong predictions, and to develop an interactive verification process for providing more finely-grained labels while minimizing expert effort.



\subsubsection{Explainable Feature Engineering}
Most existing works for improving feature quality focus on tabular or textual data from traditional analysis models.
The features of these data are naturally interpretable, which makes the feature engineering process simple.
In addition, features extracted by deep neural networks perform better than handcrafted ones~\cite{donahue2014decaf, Wang2020Visual}.
However, these deep features are hard to interpret due to the black box nature of deep neural networks, which brings several challenges for feature engineering.

Firstly, the extracted features are obtained in a data-driven process, which may poorly represent the original images/videos when the datasets are biased.
For example, given a dataset with only dark dogs and light cats, the extracted features may emphasize color and ignore other discriminating concepts, like shapes of faces and ears.
Without a clear understanding of these biased features, it is hard to correct them in a comprehensive way.
Thus, an interesting topic for future work is to utilize interactive visualization to disclose why the features are biased. 
The key challenge here is how to measure the information preserved or discarded by the extracted features and to visualize it in a comprehensible manner.

Moreover, redundancy exists in extracted deep features~\cite{ayinde2018building}.
Removing redundant features can lead to several benefits, such as reducing  storage requirements and improving generalization~\cite{chandrashekar2014survey}.
However, without a clear understanding of the exact meaning of features, it is hard to judge whether a feature is redundant.
Thus, an interesting future topic is to develop a visual analytics method to convey feature redundancy in a comprehensible way, which allows experts to explore it and remove redundant features.

\subsection{Opportunities during Model Building}
\subsubsection{Online Training Diagnosis}
Existing visual analytics tools for model diagnosis mostly work offline: the data for diagnosis is collected after the training process is finished. 
They have shown their capability for revealing the root causes of  failed training processes.
However, as modern machine learning models become more and more complex, training processes can last for days or even weeks.
Offline diagnosis severely restricts the ability of visual analytics to assist in training.
Thus, there is a significant need to develop visual analytics tools for online diagnosis of the training process so that model developers can identify  anomalies and promptly make corresponding adjustments to the process.
This can save much time in the trial-and-error model building process.
The key challenge for online diagnosis is to detect anomalies in the training process in a timely manner.
While it remains a difficult task to develop algorithms for automatically and accurately detecting anomalies in  real-time, interactive visualization promises a way to locate  potential errors in the training process.
Differing from offline diagnosis, the data of the training process will be continuously fed into the online analysis tool.
Thus, progressive visualization techniques are needed to produce meaningful visualization results of partial streaming data.
These techniques can help experts monitor online model training processes and identify possible issues rapidly.

\subsubsection{Interactive Model Refinement}
Recent works have explored the utilization of uncertainty to facilitate interactive model refinement~\cite{Yang2020Interactive,Liu2016An,ElAssady2020Semantic,Wang2016TopicPanorama}.
There are many methods to assign uncertainty scores to  model outputs (\eg\ based on confidence scores produced by classifiers), and visual hints can be used to guide users to examine  model outputs with high uncertainty.
Models uncertainty will be recomputed  after  user refinement, and users can perform iteratively until they are satisfied with the results.
Furthermore, additional information can also be leveraged  to provide users with more intelligent guidance to facilitate a fast and accurate model refinement process.
However, the room for improving interactive model refinement is still largely unexplored by researchers.
One possible direction is that since the refinement process usually requires several iterations, guidance in later iterations can be learned from users' previous interactions.
For example, in a clustering application, users may define some must-link or cannot-link constraints on some instance pairs, and such constraints can be used to instruct a model to split or merge some clusters in the intermediate result.
\jiazhi{In addition,} prior knowledge can be used to predict where refinements are needed.
For example,  model outputs may conflict with certain public or domain knowledge, especially for unsupervised models (\eg\ nonlinear matrix factorization and latent Dirichlet allocation for topic modeling), 
\jiazhi{which} should be considered  in the refinement process.
Therefore, such a knowledge-based strategy focuses on revealing unreasonable results produced by the models, allowing users to refine the models by adding constraints to them.

\subsection{Opportunities after Model Building}

\subsubsection{Understanding Multi\mengchen{-modal} Data}
\jiazhi{Existing works on content analysis have achieved great success in understanding single-modal data, such as texts, images, and videos.}
However, real-world applications often contain multi-\mengchen{modal} data, which combines several different content forms, such as text, audio, and images.
For example, a physician diagnoses a patient after considering multiple kinds of data, such as the medical record (texts), laboratory reports (tables), and CT scans (images).
When analyzing such multi-\mengchen{modal} data, in-depth relationships 
\jiazhi{between} different \mengchen{modal}s cannot be 
\jiazhi{well captured by simply combining  knowledge learned from single-modal models.}
\jiazhi{It is more promising to employ multi-\mengchen{modal} machine learning techniques and leverage their capability to disclose insights across different forms of data.}
To this end, a more powerful visual analytics system is crucial for understanding the output of such multi-\mengchen{modal} learning models.
\jiazhi{Many  machine learning models have been proposed to learn  joint representations of multi-modal data, including natural language, visual signals, and vocal signals~\cite{baltruvsaitis2018multimodal,lu2019vilbert}.}
Accordingly, an interesting future \yuanjun{direction} is how to effectively visualize learned joint representations of multi-modal data in an all-in-one manner, to facilitate the understanding \yuanjun{of} the data and their relationships. 
Various classic 
\jiazhi{multi-modal} tasks can be employed to enhance natural interactions in the field of visual analytics.
For example, in the vision-and-language scenario, the visual grounding task (identify the corresponding image area given the description) can be used to provide a natural interface to support natural-language-based image retrieval in a visual environment.

\subsubsection{Analyzing Concept Drifts}
In real-world applications, it is often assumed that the  mapping  from input data to output values (\eg\ prediction label) is static.
However, as data continues to arrive, the mapping between the input data and output values may change in unexpected ways~\cite{Lu2018survey}. 
In such a situation, a model trained on historical data may no longer work properly on new data.
This usually causes \yuanjun{noticeable performance degradation} when the application data does not match the training data. 
Such a non-stationary learning problem over time is known as concept drift.
As more and more machine learning applications directly consume streaming data, it is important to detect and analyze concept drift and minimize the resulting performance degradation~\cite{yang2020diagnosing, wang2020conceptexplorer}.
In the field of machine learning, three main research topics, have been studied: drift detection, drift understanding, and drift adaptation.
Machine learning researchers have proposed many automatic algorithms to detect and adapt to concept drift.
Although these algorithms can improve \yuanjun{the} adaptability of learning models in an uncertain environment, they only provide a numerical value to measure the  degree of drift at a given time. 
This makes it hard to understand why and where drift occurs.
If the adaptation algorithms fail to improve the model performance, the black-box
behavior of the adaptation models makes it difficult to diagnose the root cause of performance degradation.
As a result, model developers need tools that intuitively illustrate how data distributions have changed over time, which samples cause drift, and how the training samples and models can be adjusted to overcoming such drift.
This requirement naturally leads to a visual analytics paradigm where the expert interacts and collaborates in concept drift detection and adaptation algorithm by putting 
the 
human in the loop.
The major challenges here are how to (i) visually represent the evolution patterns of streaming data over time and effectively compare data distributions at different points in time, and (ii) tightly integrate such streaming data visualization with drift detection and adaptation algorithms to form an interactive and progressive analysis environment with the human in the loop.

%% file: 7-conclusion.tex
\section{Conclusions}\label{sec:conclusions}

This paper has comprehensively reviewed 
\jiazhi{recent \yuanjun{progress} and developments}
in visual analytics techniques for machine learning. 
These techniques are classified into three groups by the corresponding analysis stage: techniques 
\jiazhi{before}, during, and after model building. 
Each category is detailed by 
typical analysis tasks, and each task is illustrated by a set of representative works.
By comprehensively analyzing existing visual analytics research for machine learning, we also suggest six directions for future machine-learning-related visual analytics research, including improving data quality for weakly supervised learning and explainable feature engineering before model building, online training diagnosis and intelligent model refinement during model building, and multi-modal data understanding and concept drift analysis after model building. 
We hope this survey has \jiazhi{provided} an overview of visual analytics research for machine learning, facilitating  understanding of \yuanjun{state-of-the-art} knowledge in this area, and shedding light \yuanjun{on} future research. 